\documentclass[12pt,twoside,english]{article}
\usepackage[T1]{fontenc}
\usepackage[latin9]{inputenc}
\usepackage{geometry}
\usepackage{cite}
\usepackage{graphicx}
\usepackage{hyperref} 
\usepackage{amssymb} 

\def\beq{\begin{equation}}
\def\eeq{\end{equation}}

\def\beqn{\begin{eqnarray}}
\def\eeqn{\end{eqnarray}}
\def\ba{\begin{eqnarray}}
\def\ea{\end{eqnarray}}

\makeatletter
\usepackage{babel}
\makeatother
\date{}

\author{
{\bf Alessandro Cafarella$^{(1)}$\footnote{cafarella@inp.demokritos.gr} ,
Claudio Corian\`{o}$^{(2,3)}$\footnote{claudio.coriano@le.infn.it}} ,
{\bf Marco Guzzi$^{(2)}$\footnote{marco.guzzi@le.infn.it}}
\\[1cm]
{\normalsize $^{(1)}$Institute of Nuclear Physics, NCSR Demokritos, 15310 Athens, Greece}\\
{\normalsize $^{(2)}$Universit\`{a} del Salento and INFN Sezione di Lecce, 73100 Lecce, Italy}\\
{\normalsize $^{(3)}$Department of Physics, University of Crete, 71003 Heraklion, Greece}
\\[1cm]
\texttt{http://www.le.infn.it/candia}
}

\begin{document}

\title{\bf Precision Studies of the NNLO DGLAP Evolution at the LHC with \textsc{Candia}}
\maketitle

\begin{abstract}
We summarize the theoretical approach to the solution
of the NNLO DGLAP equations using methods based on the logarithmic
expansions in $x$-space and their implementation into the C program \textsc{Candia 1.0}\footnote{\texttt{http://www.le.infn.it/candia}}.
We present the various options implemented in the program
and discuss the different solutions.
The user can choose the order of the evolution, the type of the solution,
which can be either exact or truncated, and the evolution either with a fixed
or a varying flavor number, implemented in the varying-flavour-number scheme (VFNS).
The renormalization and factorization scale dependencies are treated separately.
In the non-singlet sector the program implements an exact NNLO solution.
\end{abstract}

\newpage

\section*{PROGRAM SUMMARY}

\begin{small}
\noindent
{\em Program Title:} \textsc{Candia}\\
{\em Journal Reference:}                                      \\
{\em Catalogue identifier:}                                   \\
{\em Licensing provisions:} none\\
{\em Programming language:} C and Fortran\\
{\em Computer:} all\\
{\em Operating system:} Linux                                       \\
{\em RAM:} In the given examples, it ranges from 4 MB to 490 MB\\
{\em Keywords:} DGLAP evolution equation, parton distribution functions, $x$-space solutions, QCD  \\
{\em PACS:} 11.10.Hi                                                  \\
{\em Classification:} 11.5  Quantum Chromodynamics, Lattice Gauge Theory, 11.1  General, High Energy
Physics and Computing.\\

\noindent
{\em Nature of problem:}\\
The program provided here solves the DGLAP evolution equations for the parton distribution functions up to NNLO.
   \\
{\em Solution method:}\\
The algorithm implemented is based on the theory of the logarithmic expansions in Bjorken-$x$ space
   \\
{\em Additional comments:}\\
In order to be sure to get the last version of the program, we suggest to download the code from our official website of \textsc{Candia} is \texttt{http://www.le.infn.it/candia}.
   \\
{\em Running time:}\\
In the given examples, it ranges from 1 to 40 minutes. The jobs have been executed on an Intel Core
2 Duo T7250 CPU at 2 GHz with a 64 bit Linux kernel. The test run script included in the
package contains 5 sample runs and may take a number of hours to process, depending on the speed of
your processor and the amount your RAM.
   \\
\end{small}

\section*{LONG WRITE UP}

\section{Introduction}

Perturbative predictions in QCD are going to be essential for the discovery of
new physics at the LHC, the new hadron collider at CERN, and for this reason the determination
of the QCD background for important processes requires the analysis of cross sections
at higher perturbative orders in an expansion in the strong coupling constant $\alpha_s$.
By now, the level of theoretical accuracy reached in the calculation of several hadronic
observables for LHC studies is rather impressive, and this has been possible thanks to the
development of new perturbative techniques which have allowed to move from previous next-to-leading-order
(NLO) determinations of several key processes, to the next order in accuracy, which is the next-to-next-to-leading order (NNLO).
These computations of the hard scatterings need to be accompanied by the corresponding
NNLO DGLAP evolution in order to be phenomenologically and theoretically consistent.
Some available codes that deal with the DGLAP evolution are \textsc{Pegasus} \cite{Vogt:2004ns},
based on the use of Mellin moments, \textsc{Qcdnum} \cite{QCDNUM} and \textsc{Hoppet}
\cite{HOPPET}, both based on a discretization of $x$ and $\mu_F^2$.

There are several issues that need to be addressed when we move to
this perturbative order. One of them concerns the size of the corrections,
which are quite small compared to the NLO case (respect to the leading order (LO) result),
while another one is their dependence on the relevant scales
(factorization, renormalization) of the process, which are arbitrary.
The sensitivity on these scales is reduced by increasing the order of the perturbative expansion.
A second point concerns the specific dependence of the result on the renormalization group
evolution (RGE) and on the way we select the solution of the corresponding equations. When a generic equation is defined in terms
of a power expansion in some parameter (coupling), its solution can be either power expanded in
the same parameter or can be
computed exactly, both approaches being legitimate options.  This indetermination should be interpreted as a {\em theoretical 
error} which needs to be quantified. In fact, solving the RGEs in one way or in another is equivalent to a summation or
to a resummation of the perturbative solution, with differences between the various
methods which start appearing beyond a certain order.

\section{Brute-force and expansion-based solutions}

It is probably convenient, in order to understand the motivations
for writing \textsc{\textsc{Candia}}, to briefly go over a classification of the ways
the DGLAP equation is solved numerically and characterize the difference between
``brute-force'' and ``expansion-based'' solutions of these equations.

A brute-force solution is obtained by simply discretizing the equation
by a finite difference method, and neglecting all the issues of perturbative accuracy
that we have just mentioned.
\textsc{Candia}, on the other hand, is based on the implementation of analytical ans\"atze  which have been shown \cite{Cafarella:2005zj} to be solutions of the evolution equations. These allow to keep 
track systematically of the logarithmic corrections which are included in the final solution and to 
control directly its logarithmic accuracy. Being the logarithms accompanied by powers of  $\alpha_s$, all 
the expanded solutions correspond to solutions of a given (and different) accuracy in the strong
coupling. In particular, the program implements also resummed expansions, such as
(\ref{NLO_ansatz}) and (\ref{NNLO_ansatz}), which involve logarithms of more complicated functions. 

Notice that in the DGLAP case -- but the same issue appears in any equation whose right-hand-side is
of a given accuracy in a parametric expansion -- the kernel is only known up to a fixed order in
$\alpha_s$, and therefore it is legitimate -- in establishing the recursive form of the ansatz -- to
decide whether or not to drop all the information coming from the higher orders, orders over which
we do not have complete control from the perturbative side and that are also present in the
recursion relations.

In this sense, it is clear that the ``brute force'' solution is just 
one of the many solutions which can be obtained by the analytical expansions. It can be reproduced 
from an expansion-based approach by extending the ansatz so to include all the logarithmic powers of the form $\alpha_s^m \ln (\alpha_s/\alpha_0)^n$, with $n, m$ integers. In practice $n$ and $m$ are finite integers, but chosen sufficiently large in the actual implementation of the numerical code.

The approach followed in our program, though entirely formulated in $x$-space, shares some of the features 
which are typical of those Mellin  methods that also rely on an analytical ansatz. In fact also in this case the exact solution is obtained recursively \cite{Ellis:1993rb}. We also remark that the Mellin method, such as the one implemented in \cite{Vogt:2004ns}, and our method overlap as for overall treatment of the logarithms, though they implement different partial summations. The proof that the various analytical ans\"atze used by us 
are solutions of the evolution equations is obtained by construction, solving the recursion relations for the unknown coefficient functions of the expansion in Mellin space \cite{Cafarella:2007tj}. 

 We have implemented two classes of solutions: the ``exact'' solutions and the ``truncated'' ones. The latter are obtained by retaining contributions to the expansion up to a certain 
power of $\alpha_s$, which can be chosen by the user. We have found that the numerical change
induced by the variation of the ansatz on some hadronic observables at NNLO is of the order of a
percent. Though this variation is indeed small if compared to the change from LO to NLO of the cross
section, 
it is comparable to its change from NLO to NNLO (quantified to be around $3\%$ on the $Z$ peak
\cite{Cafarella:2007tj}). For this reason, we think that \textsc{Candia} can be of help in the
studies of resonant 
processes where a large amount of experimental data can be collected, such as on the $Z$ peak. Here the small theoretical errors that are due either to the various ways of handling the evolution or to the changes induced by going from NLO to NNLO in the hard scatterings are far larger than the experimental statistical errors coming from the direct measurements at the LHC. In this work we 
are going to describe the basic features of our program and focus essentially on the evolution part.
More tools which may be useful for QCD partonometry and for the search of extra $Z'$ will be
released in the near
future.

\section{Summary of definitions and conventions}

We start by summarizing the notations and definitions that we will be using in the description of the program.
The general mathematical structure of the DGLAP equation is
\begin{equation}
\frac{\partial}{\partial\ln \mu_F^{2}}f(x,\mu_F^{2})=P(x,\alpha_{s}(\mu_F^{2}))\otimes
f(x,\mu_F^{2})
\label{eq:DGLAP}
\end{equation}
where the convolution product is defined by
\begin{equation}
\left[a\otimes
b\right](x)=\int_{x}^{1}\frac{\textrm{d}y}{y}a\left(\frac{x}{y}\right)b(y)=\int_{x}^{1}\frac{\textrm
{d}y}{y}a(y)b\left(\frac{x}{y}\right)
\end{equation}
and the perturbative expansion of the kernels and of the beta function
up to NNLO are respectively
\begin{equation}
P(x,\alpha_{s})=\left(\frac{\alpha_{s}}{2\pi}\right)P^{(0)}(x)+\left(\frac{\alpha_{s}}{2\pi}\right)^
{2}P^{(1)}(x)+\left(\frac{\alpha_{s}}{2\pi}\right)^{3}P^{(2)}(x)+O(\alpha_{s}^{4})
\label{eq:kernel_expansion}
\end{equation}
and
\begin{equation}
\beta(\alpha_{s})=\frac{\textrm{d}\alpha_{s}(\mu_R^{2})}{\textrm{d}\ln
\mu_R^{2}}=-\frac{\beta_{0}}{4\pi}\alpha_{s}^{2}-\frac{\beta_{1}}{16\pi^{2}}\alpha_{s}^{3}-\frac{
\beta_{2}}{64\pi^{3}}\alpha_{s}^{4}+O(\alpha_{s}^{5})
\label{eq:beta_exp}
\end{equation}
where
\begin{eqnarray}
\beta_{0} & = & \frac{11}{3}N_{C}-\frac{4}{3}T_{f} \label{eq:beta0} \\
\beta_{1} & = & \frac{34}{3}N_{C}^{2}-\frac{20}{3}N_{C}T_{f}-4C_{F}T_{f} \label{eq:beta1} \\
\beta_{2} & = &
\frac{2857}{54}N_{C}^{3}+2C_{F}^{2}T_{f}-\frac{205}{9}C_{F}N_{C}T_{f}-\frac{1415}{27}N_{C}^{2}T_{f}
+\frac{44}{9}C_{F}T_{f}^{2}+\frac{158}{27}N_{C}T_{f}^{2} \label{eq:beta2}
\end{eqnarray}
and
\begin{equation}
N_{C}=3,\qquad C_{F}=\frac{N_{C}^{2}-1}{2N_{C}}=\frac{4}{3},\qquad
T_{f}=T_{R}n_{f}=\frac{1}{2}n_{f}.
\end{equation}
$N_{C}$ is the number of colors and $n_{f}$ is the number of
active flavors, selected by the mass condition $m_{q}\leq \mu_F$, for a given factorization scale
$\mu_F$. We also denote 
with $\mu_R$ the renormalization scale.

The evolution (DGLAP) kernels are distributions whose general form is given by
\begin{equation}
P(x)=P_{1}(x)+\frac{P_{2}(x)}{(1-x)_{+}}+P_{3}\delta(1-x),\label{eq:general_kernel}
\end{equation}
with a regular part $P_{1}(x)$, a ``plus-distribution'' part
$P_{2}(x)/(1-x)_{+}$ and a delta-function term $P_{3}\delta(1-x)$.
Given a continuous and differentiable function $\alpha(x)$ defined in the $[0,1)$ interval, but
singular at $x=1$, the action of the 
plus-distribution $[\alpha(x)]_{+}$  is
defined by\begin{equation}
\int_{0}^{1}f(x)[\alpha(x)]_{+}\textrm{d}x=\int_{0}^{1}\left(f(x)-f(1)\right)\alpha(x)\textrm{d}x,
\end{equation}
where $f(x)$ is a regular test function. Alternatively, an operative
definition (that assumes full mathematical meaning only when integrated)
is the following
\begin{equation}
[\alpha(x)]_{+}=\alpha(x)-\delta(1-x)\int_{0}^{1}\alpha(y)\textrm{d}y.\label{eq:def_plus}
\end{equation}

The Parton Distribution Functions (PDFs) appear in the evolution, in general, multiplied by a factor
$x$. Eq.~(\ref{eq:DGLAP}) for $\bar{f}(x)\equiv xf(x)$ 
then reads
\begin{equation}
\frac{\partial}{\partial\ln \mu_F^{2}}\bar{f}(x,\mu_F^{2})=x\left[
P(x,\alpha_{s}(\mu_F^{2}))\otimes f(x,\mu_F^{2})\right] .
\end{equation}
We are now going to compute the convolution products that we actually evaluate in the program,
having in mind the general form 
of the kernel (\ref{eq:general_kernel}).

The treatment of the delta-function part is trivial
\begin{equation}
x\left[ P_{3}\delta(1-x)\otimes f(x)\right] =
xP_{3}\int_{x}^{1}\frac{\textrm{d}y}{y}\delta(1-y)f\left(\frac{x}{y}\right)=
P_{3}\bar{f}(x),
\label{eq:conv_3}
\end{equation}
while for the regular part we get
\begin{equation}
x\left[ P_{1}(x)\otimes
f(x)\right]
=x\int_{x}^{1}\frac{\textrm{d}y}{y}P_{1}\left(\frac{x}{y}\right)f(y)=\int_{x}^{1}\frac{\textrm{d
}y}{y}\frac{x}{y}
P_{1}\left(\frac{x}{y}\right)\bar{f}(y).
\label{eq:conv_1}
\end{equation}
Since the functions that we have to integrate are strongly varying or even singular at low $x$, to
enhance the numerical accuracy we 
introduce a new integration variable $z$ defined by $y=x^z$, and Eq.~(\ref{eq:conv_1}) is mapped
into
\begin{equation}
x\left[ P_{1}(x)\otimes f(x)\right] =
-\ln x\int_{0}^{1}\textrm{d}z\, x^{1-z}P_{1}(x^{1-z})\bar{f}(x^z).
\label{eq:conv_1_mapped}
\end{equation}

Finally, for the plus-distribution part we get, after some algebraic manipulations the relation
\begin{equation}
x\left[ \frac{P_{2}(x)}{(1-x)_{+}}\otimes f(x)\right] =
\int_{x}^{1}\textrm{d}y\frac{P_{2}(y)\bar{f}(x/y)-P_{2}(1)\bar{f}(x)}{1-y}+P_2(1)\bar{f}(x)\ln(1-x)
\label{eq:conv_2}
\end{equation}
that with the mapping $y=x^z$ becomes
\begin{equation}
x\left[ \frac{P_{2}(x)}{(1-x)_{+}}\otimes f(x)\right] =
-\ln x\int_{0}^{1}\textrm{d}z\,
x^z\frac{P_{2}(x^z)\bar{f}(x^{1-z})-P_{2}(1)\bar{f}(x)}{1-x^z}+P_2(1)\bar{f}(x)\ln(1-x).
\label{eq:conv_2_mapped}
\end{equation}
The program, being entirely developed in (Bjorken's) $x$-space uses the iteration of the convolution
in the form defined above both in the non-singlet and singlet sectors. 

\section{Non-singlet and singlet structure}

We start by defining the combinations 
\begin{equation}
q_{i}^{(\pm)}=q_{i}\pm\overline{q}_{i}
\end{equation}
\begin{equation}
q^{(\pm)}=\sum_{i=1}^{n_{f}}q_{i}^{(\pm)}.
\end{equation}
From a mathematical point of view, the distributions belonging to the non-singlet sector evolve with
a decoupled DGLAP equation of form (\ref{eq:DGLAP}), while the singlet combinations mix with the
gluon distribution. The singlet DGLAP matrix equation is

\begin{equation}
\frac{\partial}{\partial\ln \mu_F^{2}}\left( \begin{array}{c}
q^{(+)}(x,\mu_F^{2})\\
g(x,\mu_F^{2})
\end{array}\right) =\left( \begin{array}{cc}
P_{qq}(x,\alpha _{s}(\mu_F^{2})) & P_{qg}(x,\alpha _{s}(\mu_F^{2}))\\
P_{gq}(x,\alpha _{s}(\mu_F^{2})) & P_{gg}(x,\alpha _{s}(\mu_F^{2}))
\end{array}\right) \otimes \left( \begin{array}{c}
q^{(+)}(x,\mu_F^{2})\\
g(x,\mu_F^{2}),
\end{array}\right)
\end{equation}
or, in vectorial notation
\begin{equation}
\frac{\partial}{\partial\ln
\mu_F^{2}}\mathbf{s}(x,\mu_F^{2})=\mathbf{P}(x,\alpha_{s}(\mu_F^{2}))\otimes
\mathbf{s}(x,\mu_F^{2}).
\end{equation}

The general structure of the non-singlet splitting functions is given by
\begin{equation}
P_{q_{i}q_{k}}=P_{\overline{q}_{i}\overline{q}_{k}}=\delta_{ik}P_{qq}^{V}+P_{qq}^{S},
\end{equation}
\begin{equation}
P_{q_{i}\overline{q}_{k}}=P_{\overline{q}_{i}q_{k}}=\delta_{ik}P_{q\bar{q}}^{V}+P_{q\bar{q}}^{S}.
\end{equation}
where $V$ and $S$ are usually referred to as the ``valence'' and ``sea'' contributions.
This leads to three types of non-singlet distributions which evolve independently:
the flavor asymmetries
\begin{equation}
q_{NS,ik}^{(\pm)}=q_{i}^{(\pm)}-q_{k}^{(\pm)}\,,
\end{equation}
governed by the combinations
\begin{equation}
P_{NS}^{\pm}=P_{qq}^{V}\pm P_{q\bar{q}}^{V}\,,
\end{equation}
and the sum of the valence distributions of all flavors $q^{(-)}$ which 
evolves with
\begin{equation}
P_{NS}^{V}=P_{qq}^{V}-P_{q\bar{q}}^{V}+n_{f}\left(P_{qq}^{S}
-P_{q\bar{q}}^{S}\right)\equiv P_{NS}^{-}+P_{NS}^{S}.
\label{eq:PNSv}
\end{equation}
Notice that the quark-quark splitting function $P_{qq}$ can be expressed as
\begin{equation}
P_{qq}=P_{NS}^{+}+n_{f}\left(P_{qq}^{S}+P_{q\bar{q}}^{S}\right)\equiv P_{NS}^{+}+P_{ps}.
\label{eq:Pqq}
\end{equation}
with $ps$ denoting the so-called ``pure singlet'' terms. We remark that
the non-singlet contribution is the most relevant one in 
Eq.~(\ref{eq:Pqq}) at large
$x$, where the \emph{pure singlet} term $P_{ps}=P_{qq}^{S}+P_{q\bar{q}}^{S}$
is very small. At small $x$, on the other hand, the latter contribution
takes over, as $xP_{ps}$ does not vanish for $x\rightarrow0$, unlike
$xP_{NS}^{+}$. 
The gluon-quark and quark-gluon entries are given by
\begin{equation}
P_{qg}=n_{f}P_{q_{i}g},
\end{equation}
\begin{equation}
P_{gq}=P_{gq_{i}}
\end{equation}
in terms of the flavor-independent splitting functions $P_{q_{i}g}=P_{\bar{q}_{i}g}$
and $P_{gq_{i}}=P_{g\bar{q}_{i}}$. With the exception of the first order part of $P_{qg}$, neither
of the quantities $xP_{qg}$, $xP_{gq}$
and $xP_{gg}$ vanish for $x\rightarrow0$.

In the expansion in powers of $\alpha_{s}$ of the evolution equations,
the flavor-diagonal (valence) quantity $P_{qq}^{V}$ is of order $\alpha_{s}$,
while $P_{q\bar{q}}^{V}$ and the flavor-independent (sea) contributions
$P_{qq}^{S}$ and $P_{q\bar{q}}^{S}$ are of order $\alpha_{s}^{2}$.
A non-vanishing difference $P_{qq}^{S}-P_{q\bar{q}}^{S}$ is present at 
order $\alpha_s^3$.

The next step is to choose a proper basis of non-singlet distributions
that allows us to reconstruct, through linear combinations, the distribution
of each parton. The singlet evolution gives us 2 distributions,
$g$ and $q^{(+)}$, so we need to evolve $2n_{f}-1$ independent
non-singlet distributions. At NNLO we choose

\begin{enumerate}
\item $q^{(-)}$, evolving with $P_{NS}^{V}$;
\item $q_{NS,1i}^{(-)}=q_{1}^{(-)}-q_{i}^{(-)}$ (for 
$2\leq i\leq n_{f}$), evolving with $P_{NS}^{-}$;
\item $q_{NS,1i}^{(+)}=q_{1}^{(+)}-q_{i}^{(+)}$ (for 
$2\leq i\leq n_{f}$), evolving with $P_{NS}^{+}$,
\end{enumerate}
and use simple relations such as
\begin{equation}
q_{i}^{(\pm)}=\frac{1}{n_{f}}\left(q^{(\pm)}+\sum_{k=1,k\neq i}^{n_{f}}q_{NS,ik}^{(\pm)}\right)
\label{eq:comb_linNS}
\end{equation}
to perform the reconstructions of the various flavors.
Choosing $i=1$ in (\ref{eq:comb_linNS}), we compute $q_{1}^{(-)}$
from the evolved non-singlets of type 1 and 2 and $q_{1}^{(+)}$ from
the evolved singlet $q^{(+)}$ and non-singlet of type 3. Then from
the non-singlets 2 and 3 we compute respectively $q_{i}^{(-)}$ and
$q_{i}^{(+)}$ for each $i$ such that $2\leq i\leq n_{f}$, and finally
$q_{i}$ and $\bar{q}_{i}$.

Moving from NNLO to NLO things simplify, as we have $P_{qq}^{S,(1)}=P_{q\bar{q}}^{S,(1)}$.
This implies (see Eq.~(\ref{eq:PNSv})) that $P_{NS}^{V,(1)}=P_{NS}^{-,(1)}$,
i.e.~the non-singlets $q^{(-)}$ and $q_{NS,ik}^{(-)}$ evolve with
the same kernel, and the same does each linear combination thereof,
in particular $q_{i}^{(-)}$ for each flavor $i$. The basis of the $2n_{f}-1$
non-singlet distributions that we choose to evolve at NLO is

\begin{enumerate}
\item $q_{i}^{(-)}$ (for each $i\leq n_{f}$), evolving with $P_{NS}^{-,(1)}$,
\item $q_{NS,1i}^{(+)}=q_{1}^{(+)}-q_{i}^{(+)}$ (for each $i$ such that
$2\leq i\leq n_{f}$), evolving with $P_{NS}^{+,(1)}$,
\end{enumerate}
and the same we do at LO, where we have in addition $P_{NS}^{+,(0)}=P_{NS}^{-,(0)}$,
being $P_{q\bar{q}}^{V,(0)}=0$.

\section{The algorithm}

We briefly review in this section the algorithm on which \textsc{Candia} is based.
More details and a theoretical discussion can be found in our previous papers
\cite{Cafarella:2005zj,Cafarella:2007tj}.

\subsection{Non-singlet: exact solution}
\label{subsec:exact_solution}

The proof of equivalence between the logarithmic expansions implemented in \textsc{Candia} and the
alternative approach based on the use of Mellin moments, as used in reference \cite{Vogt:2004ns}, is
easily established at leading order. For this we recall the definition of the Mellin transform of a
given function $a(x)$, 
\begin{equation}
a(N)=\int_{0}^{1}a(x)x^{N-1}\textrm{d}x,\end{equation}
that maps convolution products into ordinary products\begin{equation}
\left[a\otimes b\right](N)=a(N)b(N).\end{equation}
Traditional algorithms based on Mellin space solve the equations algebraically in $N$ and then
perform a numerical inversion using 
a saddle path evaluation of the complex contour. This technique is completely bypassed in our
approach. One of the advantage of \textsc{Candia} 
is that a given initial condition for the PDFs, usually formulated in $x$-space, {\em does not} need
to be fitted to a given functional form in moment space, 
which is instead typical of a given numerical implementation of the Mellin algorithm. In fact, the
functional form in moment space in some cases 
may even not be general 
enough, and may not allow the evolution of quite singular distributions at small $x$. 
From our experience, we have found that fitting special initial conditions in $x$-space forces the
user, in codes based in Mellin space, to modify the evolution code by himself, with dubious results.
 \textsc{Candia}, by eliminating this unappealing feature
of algorithms based in Mellin space, allows any initial condition to be considered and removes the
initial numerical error due to the fit of the initial condition to the pre-assigned functional form
in moment space.

Having clarified this point, we introduce a single evolution scale $Q$,  leaving to the next
sections the discussion of the separation between 
the factorization and renormalization scales.
Switching to $\alpha_{s}$ as the independent variable, the DGLAP equation (\ref{eq:DGLAP}) is
rewritten in the form

\begin{equation}
\frac{\partial
f(N,\alpha_{s})}{\partial\alpha_{s}}=\frac{P(N,\alpha_{s})}{\beta(\alpha_{s})}f(N,\alpha_{s}).\label
{eq:DGLAP_Mellin}
\end{equation}

\subsubsection{Leading order}

Inserting in Eq.~(\ref{eq:DGLAP_Mellin}) the perturbative expansions of $P(N,\alpha_{s})$ and
$\beta(\alpha_{s})$ (Equations (\ref{eq:kernel_expansion}) and (\ref{eq:beta_exp})) arrested at
LO, we get
\begin{equation}
\frac{\partial
f(N,\alpha_{s})}{\partial\alpha_{s}}=-\frac{\left(\frac{\alpha_{s}}{2\pi}\right)P^{(0)}(N)}{\frac{
\beta_{0}}{4\pi}\alpha_{s}^{2}}f(N,\alpha_{s}),
\end{equation}
which is solved by
\begin{equation}
f(N,\alpha_s)=f(N,\alpha_{0})\left(\frac{\alpha_s}{\alpha_{0}}\right)^{-\frac{2P^{(0)}(N)}{\beta_{0}
}}=f(N,\alpha_{0})\exp\left(-\frac{2P^{(0)}(N)}{\beta_{0}}\ln\frac{\alpha_s}{\alpha_{0}}\right)
\end{equation}
where we have set $\alpha_s\equiv\alpha_s(Q^{2})$ and $\alpha_{0}\equiv\alpha_s(Q_{0}^{2})$.
Performing a Taylor expansion of the exponential we get\begin{equation}
f(N,\alpha)=f(N,\alpha_{0})\sum_{n=0}^{\infty}\left(\frac{1}{n!}\left(-\frac{2P^{(0)}(N)}{\beta_{0}}
\right)^{n}\ln^{n}\frac{\alpha}{\alpha_{0}}\right).\end{equation}
The $x$-space logarithmic ansatz, that parallels this solution is

\begin{equation}
f(x,Q^{2})=\sum_{n=0}^{\infty}\frac{A_{n}(x)}{n!}\ln^{n}\frac{\alpha_s(Q^{2})}{\alpha_s(Q_{0}^{2})},
\label{eq:LOansatz}\end{equation}
where the $A_{n}(x)$'s  are unknown functions. Setting $Q=Q_{0}$ in (\ref{eq:LOansatz})
we get the initial condition\begin{equation}
f(x,Q_{0}^{2})=A_{0}(x).\end{equation}
Inserting our ansatz (\ref{eq:LOansatz}) into the DGLAP equation
(\ref{eq:DGLAP}) and using the expansion of the kernels and of the beta function
(\ref{eq:kernel_expansion}, \ref{eq:beta_exp}) arrested at the first
term, after some algebra we derive the recursion relation
\begin{equation}
A_{n+1}(x)=-\frac{2}{\beta_{0}}\left[ P^{(0)}\otimes A_{n}\right](x).
\label{An_recurrence}
\end{equation}

In the code, the value of $\bar{A}_n(x_k)\equiv xA_n(x_k)$ for the PDF with index \texttt{i} (see
Table \ref{tab:indices}), is stored in the variable \texttt{A[i][n][k]}; $x_k$ is the $x$-grid point
stored in \texttt{X[k]}.

\subsubsection{Next-to-leading order}
Let's now move to the higher orders.

At NLO Eq.~(\ref{eq:DGLAP_Mellin}) reads
\begin{equation}
\frac{\partial
f(N,\alpha_{s})}{\partial\alpha_{s}}=-\frac{\left(\frac{\alpha_{s}}{2\pi}\right)P^{(0)}
(N)+\left(\frac{\alpha_{s}}{2\pi}\right)^{2}P^{(1)}(N)}{\frac{\beta_{0}}{4\pi}\alpha_{s}^{2}+\frac{
\beta_{1}}{16\pi^{2}}\alpha_{s}^{3}}f(N,\alpha_{s}),
\end{equation}
the solution of which is
\begin{eqnarray}
f(N,\alpha_s) & = &
f(N,\alpha_{0})\left(\frac{\alpha_s}{\alpha_{0}}\right)^{-\frac{2P^{(0)}(N)}{\beta_{0}}}\left(\frac{
4\pi\beta_{0}+\alpha_s\beta_{1}}{4\pi\beta_{0}+\alpha_{0}\beta_{1}}\right)^{\frac{2P^{(0)}(N)}{
\beta_{0}}-\frac{4P^{(1)}(N)}{\beta_{1}}}\nonumber \\
 & = & f(N,\alpha_{0})e^{a(N)L}e^{c(N)M}\nonumber \\
 & = &
f(N,\alpha_{0})\left(\sum_{n=0}^{\infty}\frac{a(N)^{n}}{n!}L^{n}\right)\left(\sum_{m=0}^{\infty}
\frac{c(N)^{m}}{m!}M^{m}\right)
\end{eqnarray}
where we have set
\begin{eqnarray}
L & = & \ln\frac{\alpha_s}{\alpha_{0}}\label{eq:L_NLO}\\
M & = &
\ln\frac{4\pi\beta_{0}+\alpha_s\beta_{1}}{4\pi\beta_{0}+\alpha_{0}\beta_{1}}\label{eq:M_NLO}\\
a(N) & = & -\frac{2P^{(0)}(N)}{\beta_{0}}\\
c(N) & = & \frac{2P^{(0)}(N)}{\beta_{0}}-\frac{4P^{(1)}(N)}{\beta_{1}}.
\end{eqnarray}
We then assume an $x$-space solution of the form
\begin{eqnarray}
f(x,Q^{2}) & = &
\left(\sum_{n=0}^{\infty}\frac{A_{n}(x)}{n!}L^{n}\right)\left(\sum_{m=0}^{\infty}\frac{C_{m}(x)}{m!}
M^{m}\right)\nonumber \\
 & = &
\sum_{s=0}^{\infty}\sum_{n=0}^{s}\frac{A_{n}(x)}{n!}\frac{C_{s-n}(x)}{(s-n)!}L^{n}M^{s-n}\nonumber
\\
 & = &
\sum_{s=0}^{\infty}\sum_{n=0}^{s}\frac{B_{n}^{s}(x)}{n!(s-n)!}L^{n}M^{s-n},\label{eq:NLOansatz}
\label{NLO_ansatz}
\end{eqnarray}
where in the first step we have re-arranged the product of the two series into
a single series with a total exponent $s=n+m$, and in the last step
we have introduced the functions\begin{equation}
B_{n}^{s}(x)=A_{n}(x)C_{s-n}(x),\qquad(n\leq s).\end{equation}

Setting $Q=Q_{0}$ in (\ref{eq:NLOansatz}) we derive the initial condition on the recursive
coefficients
\begin{equation}
f(x,Q_{0}^{2})=B_{0}^{0}(x).\end{equation}
Inserting the ansatz (\ref{eq:NLOansatz}) into the DGLAP equation
(\ref{eq:DGLAP}), using the expansions
(\ref{eq:kernel_expansion}, \ref{eq:beta_exp}) arrested at the second terms, and equating the
coefficients of $\alpha$ and
$\alpha^{2}$, we find the recursion relations
\begin{eqnarray}
B_{n}^{s}(x) & = & -\frac{2}{\beta_{0}}\left[P^{(0)}\otimes
B_{n-1}^{s-1}\right](x)\label{eq:NLO_rec_diagonale}\\
B_{n}^{s}(x) & = & -B_{n+1}^{s}(x)-\frac{4}{\beta_{1}}\left[P^{(1)}\otimes
B_{n}^{s-1}\right](x).\label{eq:NLO_rec_verticale}
\end{eqnarray}
These relations allow to compute all the coefficients $B_{n}^{s}$
$(n\leq s)$ up to a chosen $s$ starting from $B_{0}^{0}$, the value of which is
given by the initial conditions. Eq.~(\ref{eq:NLO_rec_diagonale})
allows to follow a diagonal arrow in the scheme shown in Table \ref{cap:schemaNLO};
Eq.~(\ref{eq:NLO_rec_verticale}) instead allows to compute a coefficient
once we know the coefficients at its right and over it (horizontal
and vertical arrows).
If more than one recursion relation can be used to compute a coefficient $B_n^s$, the program
follows the fastest recursion chain to determine these, i.e.~(\ref{eq:NLO_rec_diagonale}) that
involves $P^{(0)}$ instead of $P^{(1)}$.
For each $s$ the code does the following:
\begin{enumerate}
\item computes all the coefficients $B_n^{s}$ with $n\neq0$ using (\ref{eq:NLO_rec_diagonale});
\item computes the coefficient $B_0^{s}$ using (\ref{eq:NLO_rec_verticale}).
\end{enumerate}

In the code, the value of $\bar{B}^s_n(x_k)\equiv xB^s_n(x_k)$ for the PDF with index \texttt{i}
(see Table \ref{tab:indices}), is stored in the variable \texttt{B[i][s][n][k]}.
\begin{table}[h]
\begin{center}
$\begin{array}{ccccccccc}
B_{0}^{0}\\
\downarrow & \searrow\\
B_{0}^{1} & \leftarrow & B_{1}^{1}\\
\downarrow & \searrow &  & \searrow\\
B_{0}^{2} & \leftarrow & B_{1}^{2} &  & B_{2}^{2}\\
\downarrow & \searrow &  & \searrow &  & \searrow\\
B_{0}^{3} & \leftarrow & B_{1}^{3} &  & B_{2}^{3} &  & B_{3}^{3}\\
\downarrow & \searrow &  & \searrow &  & \searrow &  & \searrow\\
\ldots &  & \ldots &  & \ldots &  & \ldots &  & \ldots\end{array}$
\end{center}

\caption{Schematic representation of the procedure followed to compute each
coefficient $B_{n}^{s}$.\label{cap:schemaNLO}}
\end{table}

\subsubsection{Next-to-next-to-leading order}

At NNLO Eq.~(\ref{eq:DGLAP_Mellin}) reads
\begin{equation}
\frac{\partial
f(N,\alpha_{s})}{\partial\alpha_{s}}=-\frac{\left(\frac{\alpha_{s}}{2\pi}\right)P^{(0)}
(N)+\left(\frac{\alpha_{s}}{2\pi}\right)^{2}P^{(1)}(N)+\left(\frac{\alpha_{s}}{2\pi}\right)^{3}P^{
(2)}(N)}{\frac{\beta_{0}}{4\pi}\alpha_{s}^{2}+\frac{\beta_{1}}{16\pi^{2}}\alpha_{s}^{3}+\frac{\beta_
{2}}{64\pi^{3}}\alpha_{s}^{4}}f(N,\alpha_{s}),
\end{equation}
the solution of which is
\begin{eqnarray}
f(N,\alpha_s) & = & f(N,\alpha_{0})e^{a(N)L}e^{b(N)M}e^{d(N)T}\nonumber \\
 & = &
f(N,\alpha_{0})\left(\sum_{n=0}^{\infty}\frac{a(N)^{n}}{n!}L^{n}\right)\left(\sum_{m=0}^{\infty}
\frac{b(N)^{m}}{m!}M^{m}\right)\left(\sum_{p=0}^{\infty}\frac{d(N)^{p}}{p!}T^{p}\right)
\label{eq:NNLO_Mellin_solution}
\end{eqnarray}
where we have introduced the definitions
\begin{eqnarray}
L & = & \ln\frac{\alpha_s}{\alpha_{0}}\label{eq:L_NNLO}\\
M & = &
\ln\frac{16\pi^{2}\beta_{0}+4\pi\alpha_s\beta_{1}+\alpha_s^{2}\beta_{2}}{16\pi^{2}\beta_{0}
+4\pi\alpha_{0}\beta_{1}+\alpha_{0}^{2}\beta_{2}}\label{eq:M_NNLO}\\
T & = &
\frac{1}{\sqrt{4\beta_{0}\beta_{2}-\beta_{1}^{2}}}\arctan\frac{2\pi(\alpha_s-\alpha_{0})\sqrt{
4\beta_{0}\beta_{2}-\beta_{1}^{2}}}{2\pi(8\pi\beta_{0}+(\alpha_s+\alpha_{0})\beta_{1}
)+\alpha_s\alpha_{0}\beta_{2}}\label{eq:T_NNLO}\\
a(N) & = & -\frac{2P^{(0)}(N)}{\beta_{0}}\\
b(N) & = & \frac{P^{(0)}(N)}{\beta_{0}}-\frac{4P^{(2)}(N)}{\beta_{2}}\\
d(N) & = &
\frac{2\beta_{1}}{\beta_{0}}P^{(0)}(N)-8P^{(1)}(N)+\frac{8\beta_{1}}{\beta_{2}}P^{(2)}(N).
\end{eqnarray}
Notice that, if $4\beta_{0}\beta_{2}-\beta_{1}^{2}<0$ (it occurs
for $n_{f}=6$), $T$ has to be analytically continued
\begin{equation}
T=\frac{1}{\sqrt{\beta_{1}^{2}-4\beta_{0}\beta_{2}}}\textrm{arctanh}\frac{2\pi(\alpha_s-\alpha_{0}
)\sqrt{\beta_{1}^{2}-4\beta_{0}\beta_{2}}}{2\pi(8\pi\beta_{0}+(\alpha_s+\alpha_{0})\beta_{1}
)+\alpha_s\alpha_{0}\beta_{2}}.
\end{equation}
We then assume an $x$-space solution of the form\begin{eqnarray}
f(x,Q^{2}) & = &
\left(\sum_{n=0}^{\infty}\frac{A_{n}(x)}{n!}L^{n}\right)\left(\sum_{m=0}^{\infty}\frac{B_{m}(x)}{m!}
M^{m}\right)\left(\sum_{p=0}^{\infty}\frac{D_{p}(x)}{p!}T^{p}\right)\nonumber \\
 & = &
\sum_{s=0}^{\infty}\sum_{t=0}^{s}\sum_{n=0}^{t}\frac{A_{n}(x)}{n!}\frac{B_{t-n}(x)}{(t-n)!}\frac{D_{
s-t}(x)}{(s-t)!}L^{n}M^{t-n}T^{s-t}\nonumber \\
 & = &
\sum_{s=0}^{\infty}\sum_{t=0}^{s}\sum_{n=0}^{t}\frac{C_{t,n}^{s}(x)}{n!(t-n)!(s-t)!}L^{n}M^{t-n}T^{
s-t},\label{eq:NNLOansatz}
\label{NNLO_ansatz}
\end{eqnarray}
where in the first step we have transformed the product of three series
into a single series in the total exponent $s=n+m+p$, and we have set $t=n+m$.  In the last step we
have introduced 
the functions\begin{equation}
C_{t,n}^{s}(x)=A_{n}(x)B_{t-n}(x)D_{s-t}(x),\qquad(n\leq t\leq s).\end{equation}
Setting $Q=Q_{0}$ in (\ref{eq:NNLOansatz}) we get the initial condition\begin{equation}
f(x,Q_{0}^{2})=C_{0,0}^{0}(x).\end{equation}
Inserting the ansatz (\ref{eq:NNLOansatz}) into the DGLAP equation
(\ref{eq:DGLAP}) and using the expansions of the kernel and the beta function
(\ref{eq:kernel_expansion}, \ref{eq:beta_exp}) arrested at the third order, equating the
coefficients of $\alpha$, $\alpha^{2}$ and $\alpha^{3}$ we
find the recursion relations
\begin{eqnarray}
C_{t,n}^{s}(x) & = & -\frac{2}{\beta_{0}}\left[P^{(0)}\otimes
C_{t-1,n-1}^{s-1}\right](x)\label{eq:NNLO_rec_diagonale}\\
C_{t,n}^{s}(x) & = & -\frac{1}{2}C_{t,n+1}^{s}(x)-\frac{4}{\beta_{2}}\left[P^{(2)}\otimes
C_{t-1,n}^{s-1}\right](x)\label{eq:NNLO_rec_verticale}\\
C_{t,n}^{s}(x) & = &
-2\beta_{1}\left(C_{t+1,n}^{s}(x)+C_{t+1,n+1}^{s}(x)\right)-8\left[P^{(1)}\otimes
C_{t,n}^{s-1}\right](x).\label{eq:NNLO_rec_orizzontale}
\end{eqnarray}
Also in this case, as before, when we have to compute a given coefficient $C_{t,n}^{s}$, if more
than one recursion relation is applicable, the program follows the fastest recursion chain, i.e.~in
order (\ref{eq:NNLO_rec_diagonale}), (\ref{eq:NNLO_rec_orizzontale}) and
(\ref{eq:NNLO_rec_verticale}).
At fixed $s$ the algorithm performs the following steps:\begin{enumerate}
\item computes all the coefficients $C_{t,n}^{s}$ with $n\neq0$ using (\ref{eq:NNLO_rec_diagonale});
\item computes the coefficient $C_{s,0}^{s}$ using (\ref{eq:NNLO_rec_verticale});
\item computes the coefficient $C_{t,0}^{s}$ with $t\neq s$ using (\ref{eq:NNLO_rec_orizzontale}),
in decreasing order of $t$.
\end{enumerate}
This procedure is exemplified in the scheme shown in Table \ref{cap:schemaNNLO}
for $s=4$.

In the numerical program, the value of $\bar{C}^s_{t,n}(x_k)\equiv xC^s_{t,n}(x_k)$ for the PDF with
index \texttt{i} (see Table \ref{tab:indices}), is stored in the variable \texttt{C[i][s][t][n][k]}.

\begin{table}[h]
\begin{center}
$\begin{array}{ccccccccc}
 &  &  &  &  &  &  &  & \underline{C_{4,4}^{4}}\\
\\ &  &  &  &  &  & \underline{C_{3,3}^{4}} &  & \underline{C_{4,3}^{4}}\\
\\ &  &  &  & \underline{C_{2,2}^{4}} &  & \underline{C_{3,2}^{4}} &  & \underline{C_{4,2}^{4}}\\
\\ &  & \underline{C_{1,1}^{4}} &  & \underline{C_{2,1}^{4}} &  & \underline{C_{3,1}^{4}} &  &
\underline{C_{4,1}^{4}}\\
 & \swarrow &  & \swarrow &  & \swarrow &  & \swarrow & \downarrow\\
C_{0,0}^{4} & \longleftarrow & C_{1,0}^{4} & \longleftarrow & C_{2,0}^{4} & \longleftarrow &
C_{3,0}^{4} & \longleftarrow & C_{4,0}^{4}\end{array}$
\end{center}

\caption{Schematic representation of the procedure followed to compute each
coefficient $C_{t,n}^{s}$ for $s=4$. The underlined coefficients are
computed via Eq.~(\ref{eq:NNLO_rec_diagonale}).\label{cap:schemaNNLO}}
\end{table}

\subsection{The truncated solutions}

Besides the class of solutions of Eq.~(\ref{eq:DGLAP}) described in
Section \ref{subsec:exact_solution}, which we have called \emph{exact}, there is another important
class of solutions, that we will call {\it truncated}, which are interesting since they
correspond to $x$-space solutions which
are accurate
up to a certain order in $\alpha_s$.
While in an exact solution all the logarithmic structures
are resummed into few functions (Equations~(\ref{eq:L_NLO},\ref{eq:M_NLO}) at NLO and
(\ref{eq:L_NNLO},\ref{eq:M_NNLO},\ref{eq:T_NNLO}) at NNLO),
the truncated ones are characterized by expansions in terms of simple logarithms of
$\alpha_s/\alpha_0$, retained
up to a chosen order. We give below some details on these expansions. Notice that all the recursive
solutions built in the singlet sector - except for  the 
lowest order ones, which are exact  in any approach, either in Mellin space or in $x$-space -  are
of this type.  In this sector  we can build solutions of 
increased accuracy by using truncated solutions of higher order.
To briefly discuss these types of solutions,  we perform an expansion in $\alpha_s$ of the quantity
$P/\beta$ in the Eq. (\ref{eq:DGLAP_Mellin}) and re-arrange the DGLAP
equation (NLO or NNLO) into the form
\ba
\frac{\partial{\mathbf{f}(N,\alpha_s)}}{\partial\alpha_s}=
\frac{1}{\alpha_s}\left[\mathbf{R}_0+\alpha_s \mathbf{R}_1 +\alpha_s^2
\mathbf{R}_2 + \dots +\alpha_s^{\kappa}
\mathbf{R}_{\kappa}\right]\mathbf{f}(N,\alpha_s),\,\nonumber\\
\label{trunc_singlet}
\ea
where we have considered the expansion up to order $\kappa$ and we have defined the following
linear combinations of the $\mathbf{P}(N)$ kernels
\ba
&&\mathbf{R}_0=-\frac{2}{\beta_0}{\mathbf{P}}^{(0)}\nonumber\\
&&\mathbf{R}_1=-\frac{1}{2\pi\beta_0^2}\left[2 \beta_0
{\mathbf{P}}^{(1)}-{\mathbf{P}}^{(0)}\beta_1\right]\nonumber\\
&&\mathbf{R}_2=-\frac{1}{\pi}\left(\frac{{\mathbf{P}}^{(2)} }{2\pi\beta_0}
+\frac{\mathbf{R}_1 \beta_1}{4\beta_0}+\frac{\mathbf{R}_0\beta_2}{16\pi\beta_0}\right)\,.\nonumber\\
&&\vdots
\ea
We call Eq.~(\ref{trunc_singlet}) the {\it truncated} version of the DGLAP equation, both in the
singlet and non-singlet cases.  

The truncated equation, in Mellin space, can be solved in closed form, at least in the non-singlet
case,
obtaining a solution which is different from the exact solution of Eq.~(\ref{eq:DGLAP_Mellin})
discussed before and having the following general form
\ba
\mathbf{f}(N,\alpha_s)=\left( \frac{\alpha_s}{\alpha_0}\right)^{\mathbf{R}_0}
\textrm{exp}\left\{\mathbf{R}_1 (\alpha_s-\alpha_0)+\frac{1}{2}
\mathbf{R}_2(\alpha_s^2-\alpha_0^2)+\dots
+\frac{1}{\kappa} \mathbf{R}_{\kappa} (\alpha_s^{\kappa}-\alpha_0^{\kappa})
\right\}\mathbf{f}(N,\alpha_0)\,.\nonumber\\
\ea
Even after a truncation of the equation, the corresponding solution is still affected by higher
order terms, and can be truncated.  
To obtain the truncated version of this solution is then necessary a further Taylor expansion
around the point $(\alpha_s,\alpha_0)=(0,0)$ for the two couplings -initial and final-
\ba
\mathbf{f}(N,\alpha_s)=\left( \frac{\alpha_s}{\alpha_0}\right)^{\mathbf{R}_0}
\left[1+\mathbf{R}_1 (\alpha_s-\alpha_0)+\frac{1}{2} \mathbf{R}_2(\alpha_s^2-\alpha_0^2)+\dots
+\frac{1}{\kappa} \mathbf{R}_{\kappa}
(\alpha_s^{\kappa}-\alpha_0^{\kappa})\right]\mathbf{f}(N,\alpha_0)\,.
\ea
Again, this expression holds in both singlet and non-singlet cases, thus we can generate
{\it truncated} solutions for the parton densities.

Passing to the singlet case, which is more involved, the truncated solutions of
Eq.~(\ref{trunc_singlet}) in Mellin space can be obtained
(for a review see \cite{Cafarella:2005zj,Cafarella:2007tj,Vogt:2004ns})
by the use of the $U$-matrix ansatz.
Basically this method consists of an expansion of the general solution around the LO solution
as 
\ba
&&\mathbf{f}(N,\alpha_s)=
\left[1+\alpha_s\mathbf{U}_{1}(N)+\alpha_s^2\mathbf{U}_{2}(N)+\dots
+\alpha_s^{\kappa}\mathbf{U}_{\kappa}(N)\right]\mathbf{L}(\alpha_s,\alpha_0,N)
\nonumber\\
&&\hspace{2cm}\left[1+\alpha_0\mathbf{U}_{1}(N)+\alpha_0^2\mathbf{U}_{2}(N)+\dots
+\alpha_0^{\kappa}\mathbf{U}_{\kappa}(N)\right]^{-1}
\mathbf{f}(N,\alpha_0),\,\nonumber\\
\label{Uansatz}
\ea
where the $\mathbf{L}(\alpha_s,\alpha_0,N)$ is defined as
\ba
\mathbf{L}(\alpha_s,\alpha_0,N)=\left( \frac{\alpha_s}{\alpha_0}\right)^{\hat{R_0}}
\ea
with the $\mathbf{U}$ operators defined through a chain of recursion relations,
obtained by substituting Eq.~(\ref{Uansatz}) into the truncated equation
\ba
&&\left[\mathbf{R}_0,\mathbf{U}_1\right]=\mathbf{U}_1-\mathbf{R}_1,\nonumber\\
&&\left[\mathbf{R}_0,\mathbf{U}_2\right]=-\mathbf{R}_2 -\mathbf{R}_1\cdot\mathbf{U}_1
+2\mathbf{U}_2\nonumber\\
&&\hspace{0.8cm}\vdots
\ea
With a Taylor expansion of the second line in
Eq.~(\ref{Uansatz}) we obtain the truncated solution of order $\kappa$ for the singlet/non-singlet
case
in Mellin space. Once we have obtained such a solution, the $x$-space result is achieved
by a Mellin inversion. This and the $x$-space approach merge as we increase the order of the
expansion.  
Working directly in $x$-space, one can generate truncated solutions up to certain order $\kappa$
in a very general way by considering the following logarithmic ansatz
\begin{equation}
\mathbf{f}(x,Q^{2})=\sum_{n=0}^{\infty}
\left\{ \left[\sum_{i=0}^{\kappa}\left(\alpha_{s}(Q^{2})\right)^{i}
\frac{\mathbf{S}_{n}^{i}(x)}{n!}\right]\ln^{n}\frac{\alpha_{s}(Q^{2})}{\alpha_{s}(Q_{0}^{2})}
\right\},
\label{ktruncatedseries}
\end{equation}
which is stunningly simple.
One can demonstrate \cite{Cafarella:2005zj,Cafarella:2007tj} that inserting this expression in 
the $x$-space version of Eq.~(\ref{trunc_singlet}) the solution obtained is equivalent to that of
the $U$-matrix prescription, in the sense that the terms of the two expansions are the same as far
as the two expansions are implemented up to a sufficiently large order. In our approach the solution
is expanded in terms of logarithms of $\alpha_s/\alpha_0$ and one
controls the accuracy by inserting powers of $\alpha_s$. The coefficients $\mathbf{S}_{n}^{i}(x)$
are determined by solving the chain of recursion relations generated by inserting the logarithmic
ansatz in the DGLAP, grouping the terms proportional to the same power of $\alpha_s$ and neglecting
the terms in $\alpha_s^{\kappa+1}$ and higher powers.

We obtain the following recursion relation for the $\mathbf{S}^0$ coefficient
\begin{equation}
\mathbf{S}_{n+1}^0(x)=-\frac{2}{\beta_0}\left[\mathbf{P}^{(0)}\otimes\mathbf{S}_n^0\right](x)\,,
\label{trunc_rec0}
\end{equation}
while for $\mathbf{S}^1$ we have
\begin{eqnarray}
\mathbf{S}_{n+1}^1(x) & = &
-\frac{\beta_1}{4\pi\beta_0}\mathbf{S}_{n+1}^0(x)-\mathbf{S}_n^1(x)\nonumber\\
&& -\frac{2}{\beta_0}\left[\mathbf{P}^{(0)}\otimes\mathbf{S}_n^1\right](x)
-\frac{1}{\pi\beta_0}\left[\mathbf{P}^{(1)}\otimes\mathbf{S}_n^0\right](x)\,.
\label{trunc_rec1}
\end{eqnarray}
The recursion relations for the $\mathbf{S}^i$ coefficients $(i=2,3,\ldots,\kappa)$ are, at NLO
\begin{eqnarray}
\mathbf{S}_{n+1}^i(x) & = &
-\frac{\beta_1}{4\pi\beta_0}\mathbf{S}_{n+1}^{i-1}(x)-i\mathbf{S}_n^i(x)-(i-1)\frac{\beta_1}{
4\pi\beta_0}\mathbf{S}_{n}^{i-1}(x)\nonumber\\
&& -\frac{2}{\beta_0}\left[\mathbf{P}^{(0)}\otimes\mathbf{S}_n^i\right](x)
-\frac{1}{\pi\beta_0}\left[\mathbf{P}^{(1)}\otimes\mathbf{S}_n^{i-1}\right](x)
\label{trunc_rec_i_nlo}
\end{eqnarray}
and at NNLO
\begin{eqnarray}
\mathbf{S}_{n+1}^i(x) & = &
-\frac{\beta_1}{4\pi\beta_0}\mathbf{S}_{n+1}^{i-1}(x)-\frac{\beta_2}{16\pi^2\beta_0}\mathbf{S}_{n+1}
^{i-2}(x)\nonumber\\
&&-i\mathbf{S}_n^i(x)-(i-1)\frac{\beta_1}{4\pi\beta_0}\mathbf{S}_n^{i-1}(x)-(i-2)\frac{\beta_2}{
16\pi^2\beta_0}\mathbf{S}_n^{i-2}(x)\nonumber\\
&& -\frac{2}{\beta_0}\left[\mathbf{P}^{(0)}\otimes\mathbf{S}_n^i\right](x)
-\frac{1}{\pi\beta_0}\left[\mathbf{P}^{(1)}\otimes\mathbf{S}_n^{i-1}\right](x)
-\frac{1}{2\pi^2\beta_0}\left[\mathbf{P}^{(2)}\otimes\mathbf{S}_n^{i-2}\right](x)
\label{trunc_rec_i_nnlo}
\end{eqnarray}
These relations hold both in the non-singlet and singlet cases and can be solved either in $x$-space
or in $N$-space in terms of the initial conditions $\mathbf{f}(x,\alpha_0)=\mathbf{S}_{0}^{0}(x)$. A
further check of the overlap of the two methods is 
obtained numerically.

Since there is no closed form solution for the singlet DGLAP equation, we always use the method
described in the current section in this specific sector. On the other hand, we leave to the user
the possibility to choose the method of solution for the non-singlet equations, exact or truncated
that they can be, 
and this choice can be made at compilation time (see Section \ref{subsec:compiling}). If one chooses
the exact method implemented in \texttt{candia.c}, we need to define the $\mathbf{S}$ coefficients
of eq.~(\ref{ktruncatedseries}) only for the singlet sector:
\begin{equation}
\mathbf{S}^i_{n,singlet}=\left(
\begin{array}{c}
S^i_{n,g}\\
S^i_{n,q^{(+)}}
\end{array}
\right).
\end{equation}
In the code, the value of $\bar{S}^i_n(x_k)\equiv xS^i_n(x_k)$ for the gluon is stored in the
variable \texttt{S[i][0][n][k]} and for $q^{(+)}$ in \texttt{S[i][1][n][k]}.
If one chooses instead to solve both the singlet and the non-singlet with the truncated method
(implemented in \texttt{candia\_trunc.c}) the array \texttt{S} has to accommodate also the
non-singlet distributions, and the value of $\bar{S}^i_n(x_k)$ for the PDF with index \texttt{j}
(see Table \ref{tab:indices}) is stored in the variable \texttt{S[i][j][n][k]}.

We remark that in a previous NLO version of the implementation of the algorithm both the singlet and
the non-singlet sectors had been solved using truncated ansatz with $\kappa=1$
\cite{Cafarella:2003jr}.

\section{Renormalization scale dependence}
As we move to higher orders in the expansion of the kernels, the presence of the strong coupling
constant $\alpha_s$  allows an independent renormalization scale $\mu_R$ on the right-hand-side of
the evolution equation. 
This dependence is, in general, completely unrelated to the factorization scale.
Thus, we can formally rewrite the DGLAP equation as follows
\begin{equation}
\frac{\partial}{\partial \ln \mu_F^2}\, f(x,\mu_F^2,\mu_R^2)=
P(x,\mu_F^2,\mu_R^2) \otimes f(x,\mu_F^2,\mu_R^2)\,,
\end{equation}
where the splitting functions have acquired a $\mu_R$ dependence simply by expanding
$\alpha_s(\mu_F^2)$ in terms of
$\alpha_s(\mu_R^2)$ 
\begin{equation}
\alpha_s(\mu_R^2)=\alpha_s(\mu_F^2)-\left[-\alpha_s^2(\mu_F^2)\frac{\beta_0
L}{4\pi}
+\frac{\alpha_s^3(\mu_F^2)}{(4\pi)^2}(-\beta_0^2 L^2-\beta_1
L)\right]\,.
\end{equation}
The explicit replacements of the kernels in this new re-organization of the defining equation are
given by
\begin{eqnarray}
&& P^{(0)}(x) \to P^{(0)}(x) \nonumber \\
&& P^{(1)}(x) \to P^{(1)}(x)
-\frac{\beta_0}{2}P^{(0)}(x) L \nonumber \\
&& P^{(2)}(x) \to P^{(2)}(x)-
\beta_0 L P^{(1)}(x)
-\left(\frac{\beta_1}{4} L - \frac{\beta_0^2}{4} L^2 \right) P^{(0)}(x) \label{trasf_kernel}
\end{eqnarray}
where the logarithmic structures are identified by
\begin{equation}
L=\ln\frac{\mu_F^2}{\mu_R^2}.
\end{equation}
\textsc{Candia} allows a determination of the evolution with the two scales held separate
$\mu_F\neq\mu_R$.
In particular we have implemented the case in which $\mu_F$ and $\mu_R$ are proportional. The
proportionality factor $k_R=\mu_F^2/\mu_R^2$ is entered by the user as a command-line argument (see
Section \ref{subsec:running}).
One can easily figure out, practicing with the program, that on increasing the perturbative order
the dependence on the renormalization/factorization scale reduces.

\section{Treatment of the quark mass thresholds}

If \textsc{Candia} is executed with the variable flavor number scheme (VFNS) option (\texttt{fns}
set to 1 in the command line) and with the macro \texttt{HFT} set to 1 (that is the default value)
in the file \texttt{constants.h}, the program will implement the matching conditions with $n_f$ and
$n_f+1$ light flavors 
both for the running coupling $\alpha_s$ and for the parton distributions. The transition from the
effective theory with $n_f$ light flavors to the one with $n_f+1$ is made when the factorization
scale reaches the renormalized pole mass of a heavy quark, $\mu_F=m_{n_f+1}$. The matching condition
up to NNLO for the running coupling is \cite{Chetyrkin:1997sg}
\begin{equation}
\alpha_s^{(n_f-1)}=
\alpha_s^{(n_f)}-
\left[\alpha_s^{(n_f)}\right]^2 \frac{L}{6\pi}+
\left[\alpha_s^{(n_f)}\right]^3
\left\{\frac{1}{\pi^2}\left[\frac{L^2}{36}-\frac{19}{24}L-\frac{7}{24}\right]\right\}+
O\left(\left[\alpha_s^{(n_f)}\right]^4\right)
\label{eq:pre}
\end{equation}
or, otherwise
\begin{equation}
\alpha_s^{(n_f+1)}=
\alpha_s^{(n_f)}+
\left[\alpha_s^{(n_f)}\right]^2 \frac{L}{6\pi}+
\left[\alpha_s^{(n_f)}\right]^3 \left\{\frac{1}{48\pi^2}\left[14+38L+\frac{4}{3}L^2\right]\right\}+
O\left(\left[\alpha_s^{(n_f)}\right]^4\right),
\label{eq:post}
\end{equation}
where $L=\ln k_R=\ln(\mu_R^2/\mu_F^2)$.

The matching conditions for the parton distributions up to NNLO are \cite{Buza:1996wv}
\begin{equation}
l_i^{(n_f+1)}(x)=l_i^{(n_f)}(x)+\left(\frac{\alpha_s^{(n_f+1)}}{4\pi}\right)^2
\left[A^{NS,(2)}_{qq,H}\otimes l_i^{(n_f)}\right](x)
\label{eq:matching1}
\end{equation}
where $l$=$q,\bar{q}$ and $i=1,2,\ldots,n_f$, are the light quark/antiquark flavours;
\begin{equation}
g^{(n_f+1)}(x)=g^{(n_f)}(x)+\left(\frac{\alpha_s^{(n_f+1)}}{4\pi}\right)^2
\left\{\left[A^{S,(2)}_{gq,H}\otimes q^{(+),(n_f)}\right](x)+
\left[A^{S,(2)}_{gg,H}\otimes g^{(n_f)}\right](x)\right\}
\end{equation}
for the gluons, and
\begin{equation}
q_{n_f}^{(n_f+1)}(x)=\bar{q}_{n_f}^{(n_f+1)}(x)=\left(\frac{\alpha_s^{(n_f+1)}}{4\pi}\right)^2
\left\{\left[\tilde{A}^{S,(2)}_{Hq}\otimes q^{(+),(n_f)}\right](x)+
\left[\tilde{A}^{S,(2)}_{Hg}\otimes g^{(n_f)}\right](x)\right\}
\label{eq:matching3}
\end{equation}
for the heavy flavors.

\section{Description of the program}

\subsection{Content of the package}

The code is unpacked with the command
\begin{verbatim}
tar zxvf candia_1.0.tar.gz
\end{verbatim}
that will create the directory \texttt{candia\_1.0}, containing the following files.

\begin{itemize}

\item \texttt{candia.c} and \texttt{candia\_trunc.c} are the files including the \texttt{main}
function, each one implementing a different method of solution: the former solves the non-singlet
sector with the exact solution, while the latter uses the truncated method.

\item \texttt{makefile} and \texttt{makefile\_trunc} are the corresponding makefiles. If one is 
using other compilers than \texttt{gcc} and \texttt{gfortran} these files need to be edited.

\item \texttt{constants.h} is a header file containing some parameters that the user may want to
edit.

\item \texttt{xpns2p.f} and \texttt{xpij2p.f} are Fortran codes \cite{Moch:2004pa,Vogt:2004mw} in
which
a parametrized form of the NNLO kernels is defined. Very few modifications have been
done to make them compatible with our code.

\item \texttt{hplog.f} is a Fortran code \cite{Gehrmann:2001pz}
in which a subroutine that computes numerically the harmonic polylogarithms
up to weight 4 is implemented. Harmonic polylogarithms are defined
in \cite{Remiddi:1999ew}.

\item \texttt{partonww.f} is just a merging of the three Fortran codes \texttt{mrst2001lo.f},
\texttt{mrst2001.f} and \texttt{mrstnnlo.f} by the MRST group \cite{Martin:2001es,Martin:2002dr}
to access their grids of LO, NLO and NNLO parton densities. Very
few modifications have been done.

\item \texttt{lo2002.dat}, \texttt{alf119.dat} and \texttt{vnvalf1155.dat}
are the MRST parton densities grids at LO, NLO and NNLO respectively.

\item \texttt{a02m.f} is the Fortan code by Alekhin \cite{Alekhin:2002fv} to access his grids of LO,
NLO and NNLO parton densities.

\item \texttt{a02m.pdfs\_i\_vfn} and \texttt{a02m.dpdfs\_i\_vfn} with $i=1,2,3$
are the files in which the grids of the Alekhin parton densities in the variable flavor number
scheme are stored.

\end{itemize}

\subsection{How to run the program}

Let us describe the different steps that the user will encounter in a run of \textsc{Candia}.

\subsubsection{Editing the file \texttt{constants.h}}

The header file \texttt{constants.h} contains some macros and two arrays that the user may want to
change before compiling. The macros are described in Table \ref{tab:macros}.

\begin{table}[h]

\begin{center}
\begin{tabular}{lll}
\hline
Macro & Default & Description \\
\hline
\texttt{GRID\_PTS} & 501 & Number of points in the $x$-grid \\
\texttt{NGP} & 30 & Number of Gaussian points \\
\texttt{ITERATIONS} & 15 & Number of iterations \\
\texttt{INTERP\_PTS} & 4 & Parameter used in the polynomial interpolation (see Section
\ref{subsec:interp}) \\
\texttt{HFT} & 1 & Switch for the heavy flavors treatment \cite{Buza:1996wv,Chetyrkin:1997sg} \\
\hline
\end{tabular}
\end{center}

\caption{Macros defined in \texttt{constants.h}.}
\label{tab:macros}

\end{table}

Besides these macros, two important arrays are defined in \texttt{constants.h}.

\begin{itemize}

\item \texttt{xtab}. The values of $x$ listed in this array need to be in increasing order. The
first (lower) value is the lower value of $x$ in the grid. The last (upper) value must be 1. The
eventual intermediate values will be forced to be in the grid. This array must contain at least two
values.

\item \texttt{Qtab}. An output file with the PDF values computed at the end of the evolution will
be generated for all the values of $\mu_F$ listed in this array.

\end{itemize}

\subsubsection{Compiling}
\label{subsec:compiling}

At this point the user has to choose the solution method for the non-singlet sector.

The main file in which the exact method is implemented and the corresponding makefile are
\texttt{candia.c} and \texttt{makefile}, while the truncated method is implemented in
\texttt{candia\_trunc.c} and the corresponding makefile is \texttt{makefile\_trunc}.

In Table \ref{tab:compiling} we show for each method of solution the compilation command that the
user should type and the executable file that will be produced.

\begin{table}[h]

\begin{center}
\begin{tabular}{lll}
\hline
Method & Command & Executable \\
\hline
exact & \texttt{make} & \texttt{candia.x} \\
truncated & \texttt{make -f makefile\_trunc} & \texttt{candia\_trunc.x} \\
\hline
\end{tabular}
\end{center}

\caption{Summary of compilation commands.}
\label{tab:compiling}

\end{table}

\subsubsection{Running}
\label{subsec:running}

Let us suppose that the user wants to use the exact method of solution, so the executable is called
\texttt{candia.x}. The whole procedure applies for the truncated method as well, replacing
\texttt{candia.x} with \texttt{candia\_trunc.x}.

The user has to supply six parameters to the command line. To have a quick usage update  he/she can
just type the line 
\begin{verbatim}
candia.x
\end{verbatim}
whose output is self-explanatory.

\begin{verbatim}
USAGE

candia.x <perturbative_order> <truncation_index> <input_model> <kr> <fns> <ext>

<perturbative_order> can be 0, 1 or 2
<truncation_index> cannot be less than <perturbative_order>
<input_model>:
   0= Les Houches toy model
   1= MRST parametrization
   2= MRST grid at 1.25 GeV^2 (minimal value in their grid)
   3= Alekhin parametrization
   4= Alekhin grid at 1 GeV^2
<kr> is the ratio mu_r^2 / mu_f^2
<fns>:
   0= fixed flavor number scheme
   1= variable flavor number scheme
<ext> is the extension of the output files (max 3 characters allowed)
\end{verbatim}

For example, to run the program with the exact solution method at NNLO, with a truncation index
$\kappa=6$ and 
the MRST grid as input, with $\mu_R^2=\mu_F^2$, in the variable flavor number scheme and choosing
the extension 
\texttt{dat} for the output files, one should type the command
\begin{verbatim}
candia.x 2 6 2 1 1 dat
\end{verbatim}

\subsubsection{Understanding the output files}
For example, if we choose \texttt{dat} as an extension of the output files and our \texttt{Qtab}
array is defined by the following line in \texttt{constants.h}
\begin{verbatim}
const double Qtab[]={50.,100.,150.,200.}
\end{verbatim}
when the program exits, the user will find in the directory some files called \texttt{Qi.dat}, with
$i$
ranging from 0 to the number of elements of \texttt{Qtab} (4 in our example).
\texttt{Q0.dat} is the summary of the relevant output files, and in our example it will look like
\begin{verbatim}
file                     Q             Q^2
------------------------------------------
Q1.dat                  50            2500
Q2.dat                 100           10000
Q3.dat                 150           22500
Q4.dat                 200           40000
\end{verbatim}

The other files (for example \texttt{Q1.dat}, showing the PDFs at 50 GeV) have one line for each
point in the $x$-grid and 14 columns: $x$, $xg$, $xu$, $xd$, $xs$, $xc$, $xb$, $xt$, $x\bar{u}$,
$x\bar{d}$, $x\bar{s}$, $x\bar{c}$, $x\bar{b}$ and $x\bar{t}$.

To print the output in a different form, one should edit a small portion of the code.

\subsection{Functions}

\subsubsection{Safe allocation functions}

\begin{verbatim}
void *Malloc(size_t size);
void *Calloc(size_t nmemb,size_t size);
\end{verbatim}

These functions have the same syntax as \texttt{malloc} and \texttt{calloc} of Standard C. The only
difference is that they check if there is enough memory available to perform the allocation, and if
not they terminate the execution of the program.

\subsubsection{Function \texttt{gauleg}}

\begin{verbatim}
void gauleg(double x1,double x2,double *x,double *w,int n);
\end{verbatim}

This function is taken from \cite{press} with just minor changes.
Given the lower and upper limits of integration \texttt{x1} and \texttt{x2},
and given \texttt{n}, \texttt{gauleg} returns arrays \texttt{x{[}0,...,n-1{]}}
and \texttt{w{[}0,...,n-1{]}} of length \texttt{n}, containing the
abscissas and weights of the Gauss-Legendre \texttt{n}-point quadrature
formula.

\subsubsection{Interpolation functions}
\label{subsec:interp}

\begin{verbatim}
double interp(double *A,double x);
double polint(double *xa,double *ya,int n,double x);
\end{verbatim}

Given an array \texttt{A}, representing a function known only at the
grid points, and a number \texttt{x} in the interval $[0,1]$, \texttt{interp}
returns the polynomial interpolation of grade $\texttt{INTERP\_PTS}*2-1$ of the function at the
point \texttt{x} through an appropriate call of \texttt{polint}.
The number of points used for the interpolation are controlled by the parameter
\texttt{INTERP\_PTS}. This identifies the number of 
grid points preceding and following \texttt{x} which are globally used for the interpolation (for a
total number of $\texttt{INTERP\_PTS}*2$). If the number of 
available grid points before or after \texttt{x} is smaller than the value \texttt{INTERP\_PTS},
more points at smaller or larger values of \texttt{x} are chosen, to ensure that the total number of
used points is always $\texttt{INTERP\_PTS}*2$.
The actual interpolation is done by \texttt{polint}, a slightly modified version of the function
with the same name that appears in \cite{press}.

\subsubsection{Function \texttt{convolution}}

\begin{verbatim}
double convolution(int i,double kernel(int,double),double *A);
\end{verbatim}

Given an integer \texttt{i}, to which corresponds a grid point $x_{i}$,
a two variable function \texttt{kernel(i,x)}, representing a kernel $P(x)$,
and an array \texttt{A}, representing a function $xf(x)=\bar{f}(x)$
known at the grid points, \texttt{convolution} returns $x[P\otimes f](x)$ as the sum of the three
pieces (\ref{eq:conv_1_mapped}), (\ref{eq:conv_2_mapped}) and (\ref{eq:conv_3}).
The integrals are computed using the Gauss-Legendre technique.

\subsubsection{Function \texttt{RecRel\_A}}

\begin{verbatim}
double RecRel_A(double *A,int k,double P0(int,double));
\end{verbatim}

Given an array \texttt{A}, representing a function $A(x)$ known
at the grid points, an integer \texttt{k} (to which corresponds a
grid point $x_{k}$) and a two variable function \texttt{P0(i,x)},
representing a leading order kernel, \texttt{RecRel\_A} returns
\begin{equation}
-\frac{2}{\beta_{0}}\left[ P^{(0)}\otimes A\right](x_k)
\end{equation}
i.e.~the RHS of Eq.~(\ref{An_recurrence}) (or equivalently the RHS of each component of the
vectorial equation (\ref{trunc_rec0})).

\subsubsection{Function \texttt{RecRel\_B}}

\begin{verbatim}
double RecRel_B(double *A,double *B,int k,
                double P0(int,double),double P1(int,double));
\end{verbatim}

Given two arrays \texttt{A} and \texttt{B}, representing the functions
$A(x)$ and $B(x)$ known at the grid points, an integer \texttt{k}
(to which corresponds a grid point $x_{k}$) and two functions \texttt{P0(i,x)}
and \texttt{P1(i,x)}, representing respectively the LO and NLO part
of a kernel, \texttt{RecRel\_B} returns
\begin{equation}
-\frac{2}{\beta_0}\left[P^{(0)}\otimes B\right](x_k)
-\frac{1}{\pi\beta_0}\left[P^{(1)}\otimes A\right](x_k)
\end{equation}
i.e.~the structure of the convolution products that appear in Eqs.~(\ref{trunc_rec1}) and
(\ref{trunc_rec_i_nlo}).

When $\mu_R\neq\mu_F$, some additional terms are included to take into account the transformation
rules of the kernels (\ref{trasf_kernel}).

\subsubsection{Function \texttt{RecRel\_C}}

\begin{verbatim}
double RecRel_C(double *A,double *B,double *C,int k,double P0(int,double),
                double P1(int,double),double P2(int,double));
\end{verbatim}

Given three arrays \texttt{A}, \texttt{B} and \texttt{C} representing
the functions $A(x)$, $B(x)$ and $C(x)$ known at the
grid points, an integer \texttt{k} (to which corresponds a grid point
$x_{k}$) and three functions \texttt{P0(i,x)}, \texttt{P1(i,x)} and
\texttt{P2(i,x)}, representing respectively the LO, NLO and NNLO part
of a kernel, \texttt{RecRel\_C} returns
\begin{equation}
-\frac{2}{\beta_0}\left[P^{(0)}\otimes C\right](x_k)
-\frac{1}{\pi\beta_0}\left[P^{(1)}\otimes B\right](x_k)
-\frac{1}{2\pi^2\beta_0}\left[P^{(2)}\otimes A\right](x_k)
\end{equation}
i.e.~the structure of the convolution products that appear in Eq.~(\ref{trunc_rec_i_nnlo}).

When $\mu_R\neq\mu_F$, some additional terms are included to take into account the transformation
rules of the kernels (\ref{trasf_kernel}).

\subsubsection{Function \texttt{RecRel\_Diag}}

\begin{verbatim}
double RecRel_Diag(double *C,int k,double P0(int,double));
\end{verbatim}

The same as \texttt{RecRel\_A}.

\subsubsection{Function \texttt{RecRel\_Vert1}}

\begin{verbatim}
double RecRel_Vert1(double *B,int k,double P0(int,double),double P1(int,double));
\end{verbatim}

Given an array \texttt{B}, representing a function $B(x)$ known at the grid points, an integer
\texttt{k} (to which corresponds a grid point $x_{k}$) and two functions \texttt{P0(i,x)}
and \texttt{P1(i,x)}, representing respectively the LO and NLO part
of a kernel, \texttt{RecRel\_Vert1} returns
\begin{equation}
-\frac{4}{\beta_{1}}\left[P^{(1)}\otimes B\right](x_k)
\end{equation}
i.e.~the convolution product that appears in Eq.~(\ref{eq:NLO_rec_verticale}).

When $\mu_R\neq\mu_F$, some additional terms are included to take into account the transformation
rules of the kernels (\ref{trasf_kernel}).

\subsubsection{Function \texttt{RecRel\_Vert2}}

\begin{verbatim}
double RecRel_Vert2(double *C,int k,double P0(int,double),
                    double P1(int,double),double P2(int,double));
\end{verbatim}

Given an array \texttt{C} representing a function $C(x)$ known at the
grid points, an integer \texttt{k} (to which corresponds a grid point
$x_{k}$) and three functions \texttt{P0(i,x)}, \texttt{P1(i,x)} and
\texttt{P2(i,x)}, representing respectively the LO, NLO and NNLO part
of a kernel, \texttt{RecRel\_Vert2} returns
\begin{equation}
-\frac{4}{\beta_{2}}\left[P^{(2)}\otimes C\right](x_k)
\end{equation}
i.e.~the convolution product that appears in Eq.~(\ref{eq:NNLO_rec_verticale}).

When $\mu_R\neq\mu_F$, some additional terms are included to take into account the transformation
rules of the kernels (\ref{trasf_kernel}).

\subsubsection{Function \texttt{RecRel\_Horiz}}

\begin{verbatim}
double RecRel_Horiz(double *C,int k,double P0(int,double),double P1(int,double));
\end{verbatim}

Given an array \texttt{C}, representing a function $C(x)$ known at the grid points, an integer
\texttt{k} (to which corresponds a grid point $x_{k}$) and two functions \texttt{P0(i,x)}
and \texttt{P1(i,x)}, representing respectively the LO and NLO part
of a kernel, \texttt{RecRel\_Horiz} returns
\begin{equation}
-8\left[P^{(1)}\otimes C\right](x_k)
\end{equation}
i.e.~the convolution product that appears in Eq.~(\ref{eq:NNLO_rec_orizzontale}).

When $\mu_R\neq\mu_F$, some additional terms are included to take into account the transformation
rules of the kernels (\ref{trasf_kernel}).

\subsubsection{$\mathbf{\beta}$ function}

\begin{verbatim}
double Beta0(int f);
double Beta1(int f);
double Beta2(int f);
double Beta(int order,double alpha);
\end{verbatim}

Given an integer \texttt{f}, representing the number of active flavors $n_f$, \texttt{Beta0},
\texttt{Beta1} and \texttt{Beta2} return respectively $\beta_0$, $\beta_1$ and $\beta_2$ as in
Eqs.~(\ref{eq:beta0}), (\ref{eq:beta1}) and (\ref{eq:beta2}).

Given the perturbative \texttt{order} and the value \texttt{alpha} of $\alpha_s$, \texttt{Beta}
returns the value of the $\beta$ function given by Eq.~(\ref{eq:beta_exp}) arrested at the chosen
order.

\subsubsection{Function \texttt{alpha\_rk}}

\begin{verbatim}
double alpha_rk(int order,double alpha0,double Qi,double Qf);
\end{verbatim}

Given the perturbative \texttt{order}, an energy scale \texttt{Qi} ($Q_i$) where $\alpha_s$ has the
known value \texttt{alpha0} ($\alpha_0$) and another energy scale \texttt{Qf} ($Q_f$),
\texttt{alpha\_rk} returns $\alpha_s(Q_f)$, i.e.~the solution of the Cauchy problem
\begin{equation}
\frac{\mbox{d}\alpha_s(t)}{\mbox{d}t}=\beta({\alpha_s}),
\end{equation}
where $t=\ln(Q^2)$, with the boundary condition $\alpha_s(t_0)\equiv\alpha_s(\ln(Q_i^2))=\alpha_0$.
The differential equation has a closed form solution only at LO; at higher orders, a fourth order
Runge-Kutta method is used. To get a correct result, the number of active flavors at both energy
scales must be the same.

\subsubsection{Functions \texttt{post} and \texttt{pre}}

\begin{verbatim}
double post(double pre,int order);
double pre(double post,int order);
\end{verbatim}

These functions implement the matching condition at a quark mass threshold for $\alpha_s$, as in
Eqs.~(\ref{eq:post}) and (\ref{eq:pre}) arrested at the given perturbative \texttt{order}.
\texttt{post} returns $\alpha_s^{(n_f+1)}$ as in Eq.~(\ref{eq:post}) given $\alpha_s^{(n_f)}$
(\texttt{pre}), while \texttt{pre} returns $\alpha_s^{(n_f-1)}$ as in Eq.~(\ref{eq:pre}) given
$\alpha_s^{(n_f)}$ (\texttt{post}).

\subsubsection{Special functions}

\begin{verbatim}
double Li2(double x);
double Li3(double x);
double S12(double x);
\end{verbatim}

These functions return respectively
\begin{equation}
\mbox{Li}_2(x)=-\int^{x}_{0}\frac{\ln(1-y)}{y}\mbox{d}y
\end{equation}
\begin{equation}
\mbox{Li}_3(x)=\int^{x}_{0}\frac{\mbox{Li}_2(y)}{y}\mbox{d}y
\end{equation}
\begin{equation}
\mbox{S}_{1,2}=\frac{1}{2}\int^{x}_{0}\frac{\ln^{2}(1-y)}{y}\mbox{d}y
\end{equation}
using the program \texttt{hplog} \cite{Gehrmann:2001pz}.

\subsubsection{Function \texttt{fact}}

\begin{verbatim}
double fact(int n);
\end{verbatim}

This function returns the factorial $n!$

\subsubsection{MRST initial distributions}

\begin{verbatim}
double xuv_1(int order,double x);
double xdv_1(int order,double x);
double xS_1(int order,double x);
double xg_1(int order,double x);
double xD_1(int order,double x);
\end{verbatim}

Given the perturbative \texttt{order} and the Bjorken variable \texttt{x}, these functions return
respectively $xu_V(x)$, $xd_V(x)$, $xS(x)$, $xg(x)$ and $x\Delta(x)$ in parametric form at
$Q_0^2=1\, \mbox{GeV}^2$ as defined in \cite{Martin:2002dr} and \cite{Martin:2001es}.

\subsubsection{Alekhin initial distributions}

\begin{verbatim}
double xuv_2(int order,double x);
double xus_2(int order,double x);
double xdv_2(int order,double x);
double xds_2(int order,double x);
double xss_2(int order,double x);
double xg_2(int order,double x);
\end{verbatim}

Given the perturbative \texttt{order} and the Bjorken variable \texttt{x}, these functions return
respectively $xu_V(x)$, $xu_s(x)$, $xd_V(x)$, $xd_S(x)$, $xs_S(x)$ and $xG(x)$ in parametric form at
$Q_0^2=9\, \mbox{GeV}^2$ as defined in \cite{Alekhin:2002fv}.

\subsubsection{Initial distributions for the Les Houches toy model}

\begin{verbatim}
double xuv_3(double x);
double xdv_3(double x);
double xg_3(double x);
double xdb_3(double x);
double xub_3(double x);
double xs_3(double x);
\end{verbatim}

Given the Bjorken variable \texttt{x}, these functions return
respectively $xu_v(x)$, $xd_v(x)$, $xg(x)$, $x\bar{d}(x)$, $x\bar{u}(x)$ and $xs(x)$ in parametric
form at $Q_0^2=2\, \mbox{GeV}^2$ as defined in \cite{Giele:2002hx}.

\subsubsection{LO kernels}
\label{sec:LOkernels}

\begin{verbatim}
double P0NS(int i,double x);
double P0qq(int i,double x);
double P0qg(int i,double x);
double P0gq(int i,double x);
double P0gg(int i,double x);
\end{verbatim}

Given the Bjorken variable \texttt{x}, these functions return, depending
on the value of the index \texttt{i}
\begin{enumerate}
\item the regular part of the kernel $P_{1}(x)$;
\item the plus-distribution part of the kernel $P_{2}(x)$;
\item the delta-function part of the kernel $P_{3}$,
\end{enumerate}
as in Eq.~(\ref{eq:general_kernel}). The kernels returned are respectively $P^{(0)}_{NS}$,
$P^{(0)}_{qq}$, $P^{(0)}_{qg}$, $P^{(0)}_{gq}$ and $P^{(0)}_{gg}$ \cite{Furmanski:1981ja}.

\subsubsection{NLO kernels}

\begin{verbatim}
double P1NSm(int i,double x);
double P1NSp(int i,double x);
double P1qq(int i,double x);
double P1qg(int i,double x);
double P1gq(int i,double x);
double P1gg(int i,double x);
\end{verbatim}

Given the Bjorken variable \texttt{x} and the index \texttt{i} with the same meaning of Section
\ref{sec:LOkernels}, these functions return respectively $P^{-,(1)}_{NS}$, $P^{+,(1)}_{NS}$,
$P^{(1)}_{qq}$, $P^{(1)}_{qg}$, $P^{(1)}_{gq}$ and $P^{(1)}_{gg}$ \cite{Furmanski:1981ja}.

\subsubsection{NNLO kernels}

\begin{verbatim}
double P2NSm(int i,double x);
double P2NSp(int i,double x);
double P2NSv(int i,double x);
double P2qq(int i,double x);
double P2qg(int i,double x);
double P2gq(int i,double x);
double P2gg(int i,double x);
\end{verbatim}

Given the Bjorken variable \texttt{x} and the index \texttt{i} with the same meaning of Section
\ref{sec:LOkernels}, these functions return respectively $P^{-,(2)}_{NS}$, $P^{+,(2)}_{NS}$,
$P^{V,(2)}_{NS}$, $P^{(2)}_{qq}$, $P^{(2)}_{qg}$, $P^{(2)}_{gq}$ and $P^{(2)}_{gg}$ in a
parametrized form \cite{Moch:2004pa,Vogt:2004mw}, using the Fortran files \texttt{xpns2p.f} and
\texttt{xpij2p.f}.

\subsubsection{Quark mass threshold kernels}

\begin{verbatim}
double A2ns(int i,double z);
double A2gq(int i,double z);
double A2gg(int i,double z);
double A2hq(int i,double z);
double A2hg(int i,double z);
\end{verbatim}

Given the Bjorken variable \texttt{z} and the index \texttt{i} with the same meaning of Section
\ref{sec:LOkernels}, these functions return respectively $A^{NS,(2)}_{qq,H}$, $A^{S,(2)}_{gq,H}$,
$A^{S,(2)}_{gg,H}$, $\tilde{A}^{S,(2)}_{Hq}$ and $\tilde{A}^{S,(2)}_{Hg}$ \cite{Buza:1996wv}.

\subsection{Distribution indices and identifiers}
\label{subsec:dist_ind}

In Table \ref{tab:indices} we show the numerical index associated to each PDF appearing in the
program.
To the PDF labelled by the number \texttt{index} we associate a string \texttt{id[index]}, reported
in the third column. Although not used in the current implementation, these identifiers can be
useful if one wish to modify the output of the program: for example if one wants to print a file for
each PDF, these strings can be used as part of the file name.

\begin{table}[h]
\begin{tabular}{lll}
\hline 
0&
gluons, $g$&
\texttt{g}\tabularnewline
1-6&
quarks, $q_{i}$, sorted by increasing mass ($u,d,s,c,b,t$)&
\texttt{u,d,s,c,b,t}\tabularnewline
7-12&
antiquarks, $\overline{q}_i$&
\texttt{au,ad,as,ac,ab,at}\tabularnewline
13-18&
$q_{i}^{(-)}$&
\texttt{um,dm,sm,cm,bm,tm}\tabularnewline
19-24&
$q_{i}^{(+)}$&
\texttt{up,dp,sp,cp,bp,tp}\tabularnewline
25&
$q^{(-)}$&
\texttt{qm}\tabularnewline
26-30&
$q_{NS,1i}^{(-)},\quad i\neq1$&
\texttt{dd,sd,cd,bd,td}\tabularnewline
31&
$q^{(+)}$&
\texttt{qp}\tabularnewline
32-36&
$q_{NS,1i}^{(+)},\quad i\neq1$&
\texttt{ds,ss,cs,bs,ts}\tabularnewline
\hline
\end{tabular}
\caption{Correspondence between indices, parton distributions and identifiers.}
\label{tab:indices}
\end{table}

\subsection{Main variables}

In Table \ref{tab:variables} we describe the main variables defined in the program.

\begin{table}[h]
\begin{tabular}{ll}
\hline 
\texttt{X{[}i{]}}&
$i$-th grid point, $x_{i}$\tabularnewline
\texttt{XG{[}i{]}}&
$i$-th Gaussian abscissa in the range $[0,1]$, $X_{i}$\tabularnewline
\texttt{WG{[}i{]}}&
$i$-th Gaussian weight in the range $[0,1]$, $W_{i}$\tabularnewline
\texttt{order}&
perturbative order\tabularnewline
\texttt{trunc}&
truncation index $\kappa$\tabularnewline
\texttt{input}&
identifier of the input model (see Section \ref{subsec:running})\tabularnewline
\texttt{kr}&
$k_R=\mu_R^2/\mu_F^2$\tabularnewline
\texttt{ntab[i]}&
position in the array \texttt{X} of the $x$ value \texttt{xtab[i]}
(i.e.~\texttt{X[ntab[i]]}$=$\texttt{xtab[i]})\tabularnewline
\texttt{nf, Nf}&
number of active flavors, $n_{f}$\tabularnewline
\texttt{nfi}&
number of active flavors at the input scale\tabularnewline
\texttt{nff}&
number of active flavors in the last evolution step\tabularnewline
\texttt{beta0}&
$\beta_{0}$\tabularnewline
\texttt{beta1}&
$\beta_{1}$\tabularnewline
\texttt{beta2}&
$\beta_{2}$\tabularnewline
\texttt{beta}&
$\beta$\tabularnewline
\texttt{log\_mf2\_mr2}&
$\ln \mu_F^2/\mu_R^2$\tabularnewline
\texttt{Q{[}i{]}}&
values of $\mu_F$ between which an evolution step is performed\tabularnewline
\texttt{alpha\_pre[i]}&
value of $\alpha_s$ just below the $i$-th flavor threshold\tabularnewline
\texttt{alpha\_post[i]}&
value of $\alpha_s$ just above the $i$-th flavor threshold\tabularnewline
\texttt{alpha1}&
$\alpha_{s}(\mu_R^{in})$, where $\mu_R^{in}$ is the lower $\mu_R$ of the evolution
step\tabularnewline
\texttt{alpha2}&
$\alpha_{s}(\mu_R^{fin})$, where $\mu_R^{fin}$ is the higher $\mu_R$ of the evolution
step\tabularnewline
\texttt{A[i][n][k]}&
coefficient $\bar{A}_n(x_{k})$ for the distribution with index \texttt{i}\tabularnewline
\texttt{B[i][s][n][k]}&
coefficient $\bar{B}^s_n(x_{k})$ for the distribution with index \texttt{i}\tabularnewline
\texttt{C[i][s][t][n][k]}&
coefficient $\bar{C}^s_{t,n}(x_{k})$ for the distribution with index \texttt{i}\tabularnewline
\texttt{S[i][j][n][k]}&
\texttt{candia.c}: coefficient $\bar{S}^i_n(x_{k})$ for $g$ (\texttt{j}$=0$) or $q^{(+)}$ (\texttt{j}$=1$) \tabularnewline
&
\texttt{candia\_trunc.c}: coefficient $\bar{S}^i_n(x_{k})$ for the distribution with index \texttt{j}\tabularnewline
\texttt{R[i][k]}&
value of the PDF $\bar{f}(x_{k})$ with index \texttt{i}\tabularnewline
\hline
\end{tabular}
\caption{Main variables defined in the program.}
\label{tab:variables}
\end{table}

\section{Results}

In Tables \ref{tab:EXvsPEG} and \ref{tab:TRvsPEG} we compare the results obtained running \textsc{Candia} with the exact and the truncated method, respectively, with those obtained using \textsc{Pegasus} \cite{Vogt:2004ns} in the variable flavor number scheme and at a final scale $\mu_F=\mu_R=100\,\mbox{GeV}$.
The initial conditions are taken from the Les Houches toy model with $Q_0^2=2\,\mbox{GeV}^2$
\cite{Giele:2002hx}.
We have made these runs using the following \texttt{constants.h} file
\begin{verbatim}
#define GRID_PTS 801
#define NGP 30
#define ITERATIONS 15
#define INTERP_PTS 4
#define HFT 1

const double xtab[]={1e-7,1e-6,1e-5,1e-4,1e-3,1e-2,0.1,0.3,0.5,0.7,0.9,1.};
const double Qtab[]={100.};
\end{verbatim}
We have also chosen a truncation index $\kappa=6$ at NLO and $\kappa=7$ at NNLO.
For example, the command to run the NLO evolution with the truncated method, in this case, is 
\begin{verbatim}
candia_trunc.x 1 6 0 1 1 t1
\end{verbatim}
while that for the NNLO evolution with the exact method is
\begin{verbatim}
candia.x 2 7 0 1 1 e2
\end{verbatim}

We have chosen two representative PDFs: a non-singlet, $xq^{(-)}$, and a singlet, $xg$.

\begin{table}[h]
\begin{footnotesize}
\begin{center}
\begin{tabular}{|c|rr|rr|rr|}

\hline 
 & \multicolumn{2}{c|}{LO} & \multicolumn{2}{c|}{NLO} & \multicolumn{2}{c|}{NNLO}\\
$x$ & \multicolumn{1}{c}{$xq^{(-)}$} & \multicolumn{1}{c|}{$xg$} & \multicolumn{1}{c}{$xq^{(-)}$} & \multicolumn{1}{c|}{$xg$} & \multicolumn{1}{c}{$xq^{(-)}$} & \multicolumn{1}{c|}{$xg$}\\
\hline
$10^{-7}$ & $9.3734\cdot10^{-5}$ & $1.3272\cdot10^{+3}$ & $1.7347\cdot10^{-4}$ & $1.1203\cdot10^{+3}$ & $2.7814\cdot10^{-4}$ & $9.9238\cdot10^{+2}$\\
 & $+0.0001\cdot10^{-5}$ & $0.0000\cdot10^{+3}$ & $+0.0008\cdot10^{-4}$ & $+0.0036\cdot10^{+3}$ & $-0.6552\cdot10^{-4}$ & $-0.0456\cdot10^{+2}$\\
\hline
$10^{-6}$ & $5.4063\cdot10^{-4}$ & $6.0117\cdot10^{+2}$ & $8.8066\cdot10^{-4}$ & $5.2431\cdot10^{+2}$ & $1.2137\cdot10^{-3}$ & $4.8602\cdot10^{+2}$\\
 & $0.0000\cdot10^{-4}$ & $0.0000\cdot10^{+2}$ & $+0.0035\cdot10^{-4}$ & $+0.0141\cdot10^{+2}$ & $-0.1927\cdot10^{-3}$ & $-0.0215\cdot10^{+2}$\\
\hline
$10^{-5}$ & $3.0236\cdot10^{-3}$ & $2.5282\cdot10^{+2}$ & $4.3423\cdot10^{-3}$ & $2.2804\cdot10^{+2}$ & $5.2399\cdot10^{-3}$ & $2.1922\cdot10^{+2}$\\
 & $0.0000\cdot10^{-3}$ & $0.0000\cdot10^{+2}$ & $+0.0015\cdot10^{-3}$ & $+0.0051\cdot10^{+2}$ & $-0.4555\cdot10^{-3}$ & $-0.0090\cdot10^{+2}$\\
\hline
$10^{-4}$ & $1.6168\cdot10^{-2}$ & $9.6048\cdot10^{+1}$ & $2.0611\cdot10^{-2}$ & $8.9671\cdot10^{+1}$ & $2.2485\cdot10^{-2}$ & $8.8486\cdot10^{+1}$\\
 & $0.0000\cdot10^{-2}$ & $0.0000\cdot10^{+1}$ & $+0.0006\cdot10^{-2}$ & $+0.0158\cdot10^{+1}$ & $-0.0695\cdot10^{-2}$ & $-0.0318\cdot10^{+1}$\\
\hline
$10^{-3}$ & $8.0470\cdot10^{-2}$ & $3.1333\cdot10^{+1}$ & $9.2179\cdot10^{-2}$ & $3.0283\cdot10^{+1}$ & $9.4766\cdot10^{-2}$ & $3.0319\cdot10^{+1}$\\
 & $0.0000\cdot10^{-2}$ & $0.0000\cdot10^{+1}$ & $+0.0020\cdot10^{-2}$ & $+0.0038\cdot10^{+1}$ & $-0.0112\cdot10^{-2}$ & $-0.0085\cdot10^{+1}$\\
\hline
$10^{-2}$ & $3.4576\cdot10^{-1}$ & $7.7728\cdot10^{+0}$ & $3.6111\cdot10^{-1}$ & $7.7546\cdot10^{+0}$ & $3.6243\cdot10^{-1}$ & $7.7785\cdot10^{+0}$\\
 & $0.0000\cdot10^{-1}$ & $0.0000\cdot10^{+0}$ & $+0.0005\cdot10^{-1}$ & $+0.0055\cdot10^{+0}$ & $+0.0138\cdot10^{-1}$ & $-0.0127\cdot10^{+0}$\\
\hline
$0.1$ & $8.5499\cdot10^{-1}$ & $8.4358\cdot10^{-1}$ & $8.2582\cdot10^{-1}$ & $8.5589\cdot10^{-1}$ & $8.1971\cdot10^{-1}$ & $8.5284\cdot10^{-1}$\\
 & $0.0000\cdot10^{-1}$ & $0.0000\cdot10^{-1}$ & $+0.0006\cdot10^{-1}$ & $+0.0003\cdot10^{-1}$ & $+0.0016\cdot10^{-1}$ & $+0.0018\cdot10^{-1}$\\
\hline
$0.3$ & $5.1640\cdot10^{-1}$ & $7.8026\cdot10^{-2}$ & $4.8177\cdot10^{-1}$ & $7.9588\cdot10^{-2}$ & $4.7464\cdot10^{-1}$ & $7.9004\cdot10^{-2}$\\
 & $0.0000\cdot10^{-1}$ & $0.0000\cdot10^{-2}$ & $+0.0002\cdot10^{-1}$ & $-0.0037\cdot10^{-2}$ & $+0.0016\cdot10^{-1}$ & $+0.0106\cdot10^{-2}$\\
\hline
$0.5$ & $1.6764\cdot10^{-1}$ & $7.4719\cdot10^{-3}$ & $1.5287\cdot10^{-1}$ & $7.7199\cdot10^{-3}$ & $1.4968\cdot10^{-1}$ & $7.6537\cdot10^{-3}$\\
 & $0.0000\cdot10^{-1}$ & $0.0000\cdot10^{-3}$ & $0.0000\cdot10^{-1}$ & $-0.0066\cdot10^{-3}$ & $+0.0019\cdot10^{-1}$ & $+0.0139\cdot10^{-3}$\\
\hline
$0.7$ & $2.6157\cdot10^{-2}$ & $3.5241\cdot10^{-4}$ & $2.3196\cdot10^{-2}$ & $3.7528\cdot10^{-4}$ & $2.2536\cdot10^{-2}$ & $3.7161\cdot10^{-4}$\\
 & $0.0000\cdot10^{-2}$ & $0.0000\cdot10^{-4}$ & $0.0000\cdot10^{-2}$ & $-0.0046\cdot10^{-4}$ & $+0.0060\cdot10^{-2}$ & $+0.0081\cdot10^{-4}$\\
\hline
$0.9$ & $4.4201\cdot10^{-4}$ & $1.0308\cdot10^{-6}$ & $3.7019\cdot10^{-4}$ & $1.1939\cdot10^{-6}$ & $3.5398\cdot10^{-4}$ & $1.1764\cdot10^{-6}$\\
 & $+0.0001\cdot10^{-4}$ & $+0.0002\cdot10^{-6}$ & $+0.0001\cdot10^{-4}$ & $-0.0016\cdot10^{-6}$ & $+0.0183\cdot10^{-4}$ & $+0.0042\cdot10^{-6}$\\
\hline

\end{tabular}
\end{center}
\end{footnotesize}
\caption{Comparison between \textsc{Candia} (exact method) and \textsc{Pegasus} at $\mu_F=\mu_R=100\,\mbox{GeV}$ in the variable flavor number scheme. In each entry, the first number is \textsc{Candia}'s result and the second one is the difference between \textsc{Candia}'s and \textsc{Pegasus}' result. The Les Houches toy model at $2\,\mbox{GeV}^2$ has been used as initial condition.}
\label{tab:EXvsPEG}
\end{table}

\begin{table}[h]
\begin{footnotesize}
\begin{center}
\begin{tabular}{|c|r|rr|}

\hline 
 & \multicolumn{1}{c|}{NLO} & \multicolumn{2}{c|}{NNLO}\\
$x$ & \multicolumn{1}{c|}{$xq^{(-)}$} & \multicolumn{1}{c}{$xq^{(-)}$} & \multicolumn{1}{c|}{$xg$}\\
\hline
$10^{-7}$ & $1.6643\cdot10^{-4}$ & $3.3624\cdot10^{-4}$ & $9.9220\cdot10^{+2}$\\
 & $-0.0696\cdot10^{-4}$ & $-0.0742\cdot10^{-4}$ & $-0.0474\cdot10^{+2}$\\
\hline
$10^{-6}$ & $8.6037\cdot10^{-4}$ & $1.3840\cdot10^{-3}$ & $4.8595\cdot10^{+2}$\\
 & $-0.1994\cdot10^{-4}$ & $-0.0224\cdot10^{-3}$ & $-0.0222\cdot10^{+2}$\\
\hline
$10^{-5}$ & $4.3192\cdot10^{-3}$ & $5.6583\cdot10^{-3}$ & $2.1919\cdot10^{+2}$\\
 & $-0.0216\cdot10^{-3}$ & $-0.0371\cdot10^{-3}$ & $-0.0093\cdot10^{+2}$\\
\hline
$10^{-4}$ & $2.0826\cdot10^{-2}$ & $2.3325\cdot10^{-2}$ & $8.8478\cdot10^{+1}$\\
 & $+0.0221\cdot10^{-2}$ & $+0.0145\cdot10^{-2}$ & $-0.0326\cdot10^{+1}$\\
\hline
$10^{-3}$ & $9.4074\cdot10^{-2}$ & $9.6537\cdot10^{-2}$ & $3.0317\cdot10^{+1}$\\
 & $+0.1915\cdot10^{-2}$ & $+0.1659\cdot10^{-2}$ & $-0.0087\cdot10^{+1}$\\
\hline
$10^{-2}$ & $3.6821\cdot10^{-1}$ & $3.6771\cdot10^{-1}$ & $7.7784\cdot10^{+0}$\\
 & $+0.0715\cdot10^{-1}$ & $+0.0666\cdot10^{-1}$ & $-0.0128\cdot10^{+0}$\\
\hline
$0.1$ & $8.2161\cdot10^{-1}$ & $8.1594\cdot10^{-1}$ & $8.5290\cdot10^{-1}$\\
 & $-0.0415\cdot10^{-1}$ & $-0.0361\cdot10^{-1}$ & $+0.0024\cdot10^{-1}$\\
\hline
$0.3$ & $4.6358\cdot10^{-1}$ & $4.5766\cdot10^{-1}$ & $7.9012\cdot10^{-2}$\\
 & $-0.1817\cdot10^{-1}$ & $-0.1682\cdot10^{-1}$ & $+0.0114\cdot10^{-2}$\\
\hline
$0.5$ & $1.4288\cdot10^{-1}$ & $1.4025\cdot10^{-1}$ & $7.6545\cdot10^{-3}$\\
 & $-0.0999\cdot10^{-1}$ & $-0.0924\cdot10^{-1}$ & $+0.0147\cdot10^{-3}$\\
\hline
$0.7$ & $2.0960\cdot10^{-2}$ & $2.0420\cdot10^{-2}$ & $3.7165\cdot10^{-4}$\\
 & $-0.2236\cdot10^{-2}$ & $-0.2056\cdot10^{-2}$ & $+0.0085\cdot10^{-4}$\\
\hline
$0.9$ & $3.1502\cdot10^{-4}$ & $3.0217\cdot10^{-4}$ & $1.1765\cdot10^{-6}$\\
 & $-0.5516\cdot10^{-4}$ & $-0.4998\cdot10^{-4}$ & $+0.0043\cdot10^{-6}$\\
\hline

\end{tabular}
\end{center}
\end{footnotesize}
\caption{Same as Table \ref{tab:EXvsPEG}, but this time \textsc{Candia} has been run with the truncated method. To avoid repetitions, the columns for which the exact and the truncated methods give the same results are not shown.}
\label{tab:TRvsPEG}
\end{table}

The small difference between the NNLO gluon distributions calculated with the exact and the
truncated methods are due to the heavy quark matching conditions
(\ref{eq:matching1}-\ref{eq:matching3}).
Since a solution in closed form is not available for the singlet sector, both the exact and
truncated methods solve the singlet equations in the same way. By the way, Equations
(\ref{eq:matching1}) and (\ref{eq:matching3}), whose first-non zero contribution is of order
$\alpha_s^2$, mix the non-singlet and singlet sectors at NNLO.

We will therefore concentrate on the differences in the non-singlet sector.
At NLO it is evident that there is a better agreement for $xq^{(-)}$ with \textsc{Pegasus} using the
exact method (Table \ref{tab:EXvsPEG}). At NNLO the truncated method agrees better with
\textsc{Pegasus} at low $x$ while the exact method agrees better at high $x$, where the two regions
can be approximately separated at a value of $x$ in the interval $[10^{-4},10^{-3}]$.
This peculiar behavior is not surprising, because our exact method of solution for the non-singlet at NNLO is based on the closed form solution of the 
DGLAP equation in Mellin-space (\ref{eq:NNLO_Mellin_solution}), reconstructed via iteration in $x$-space, as we have explained before, and which is not present in \textsc{Pegasus}.

\begin{figure}[h]
\begin{center}
\includegraphics[width=9cm,angle=-90]{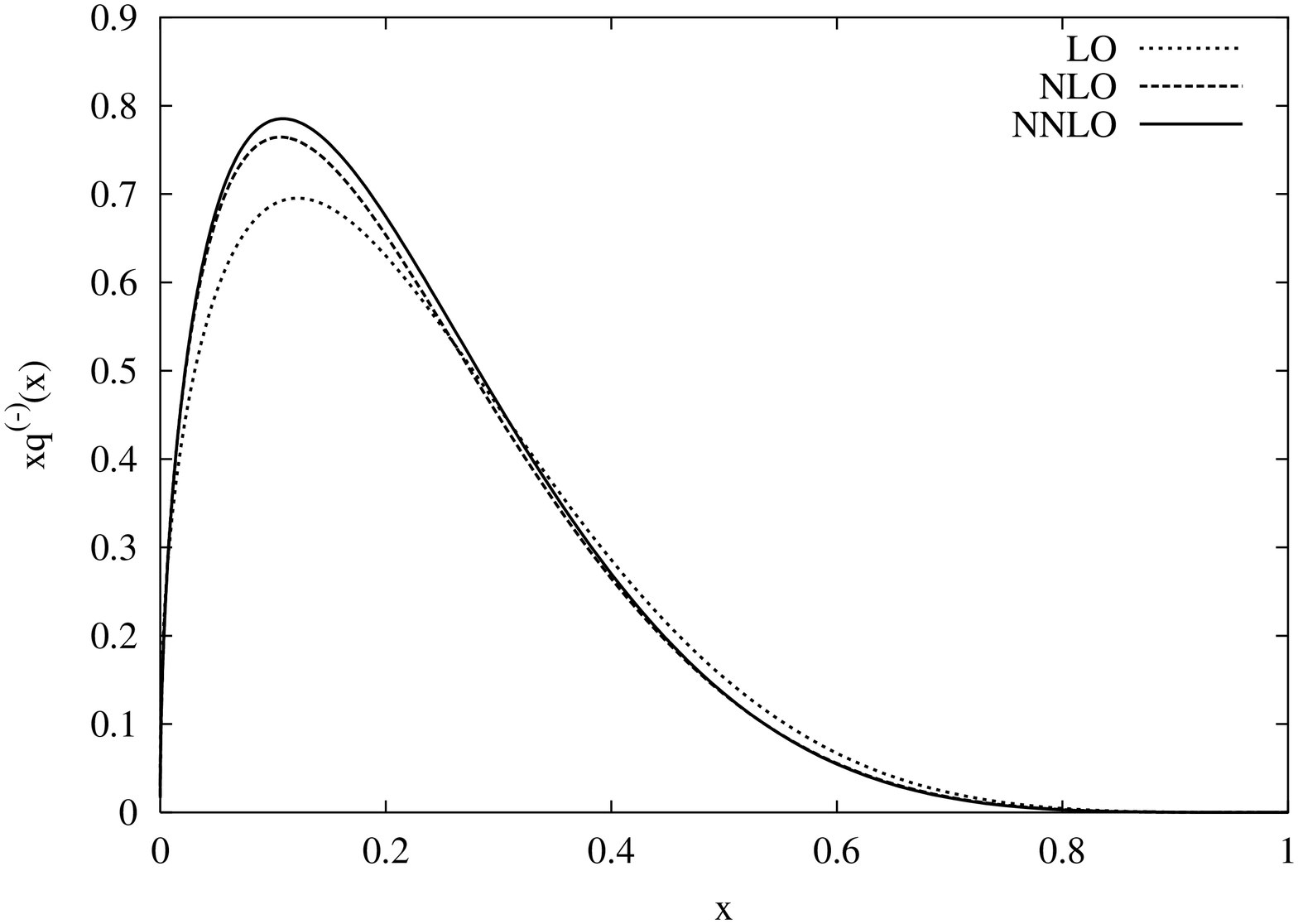}
\includegraphics[width=9cm,angle=-90]{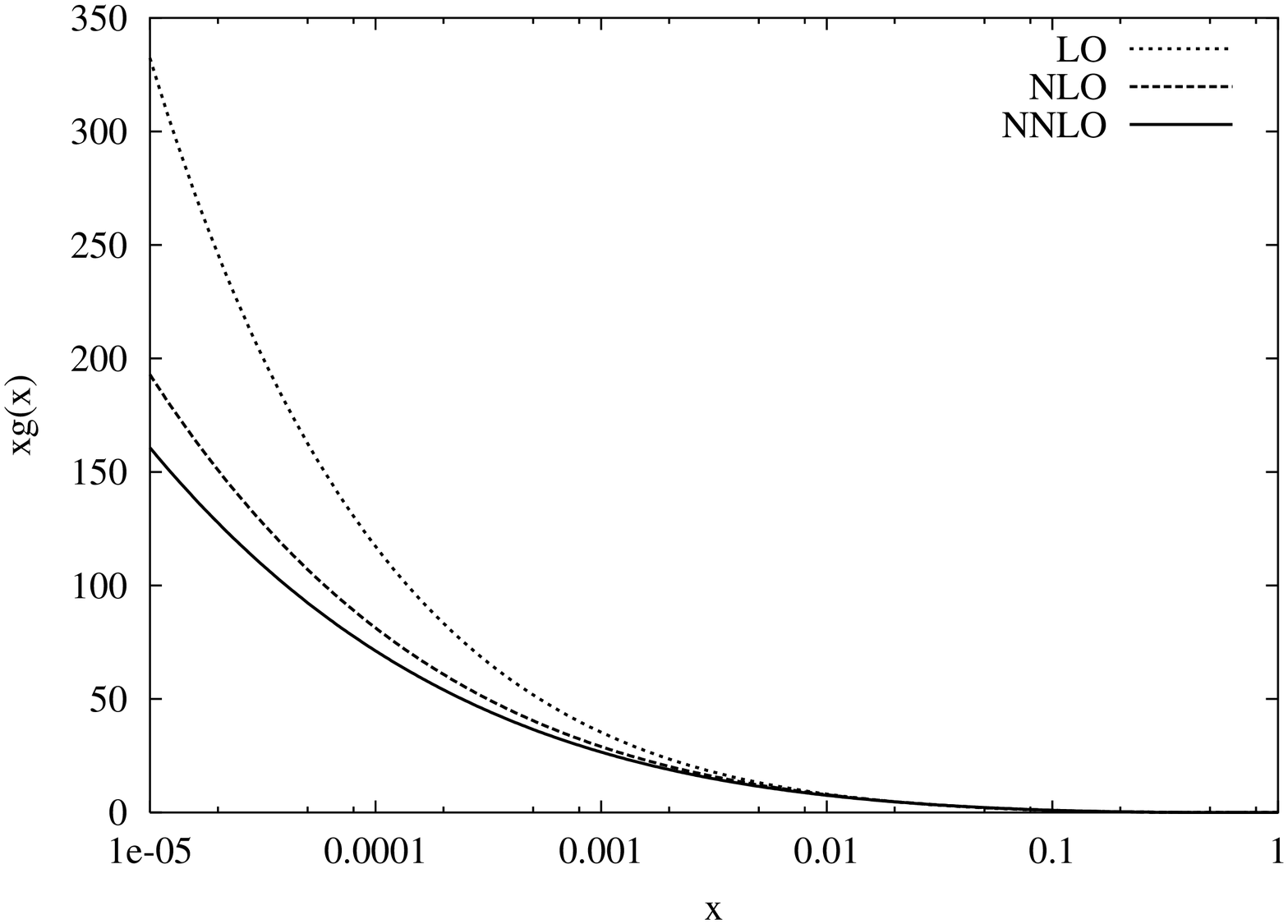}
\end{center}
\caption{Plots of $xq^{(-)}$ and $xg$ at different perturbative orders for \textsc{Candia} (exact method) at $\mu_F=\mu_R=100\,\mbox{GeV}$ in the variable flavor number scheme. The MRST parametrizations at $1\,\mbox{GeV}$ have been used as initial conditions.}
\label{fig:mrst}
\end{figure}

In Table \ref{tab:EXvsTR}, where we show the variation of $xq^{(-)}$ and $xg$ with the perturbative order, we compare the results obtained running \textsc{Candia} with the exact versus the truncated method, respectively,  
having chosen MRST initial conditions. The plots in Figure \ref{fig:mrst}, have been obtained evolving the MRST initial conditions from $1$ to $100\,\mbox{GeV}$ with \textsc{Candia} (exact method). Notice that as we move to NNLO from the LO result, the gluon density, in particular, is drastically reduced in the low-$x$ region.
For the MRST runs we have modified the file \texttt{constants.h} as follows
\begin{verbatim}
#define GRID_PTS 501
#define NGP 30
#define ITERATIONS 15
#define INTERP_PTS 4
#define HFT 1

const double xtab[]={1e-5,1e-4,1e-3,1e-2,0.1,0.2,0.3,0.4,0.5,0.6,0.7,0.8,0.9,1.};
const double Qtab[]={100.};
\end{verbatim}

\begin{table}[h]
\begin{footnotesize}
\begin{center}
\begin{tabular}{|c|r|rr|}

\hline 
 & \multicolumn{1}{c|}{NLO} & \multicolumn{2}{c|}{NNLO}\\
$x$ & \multicolumn{1}{c|}{$xq^{(-)}$} & \multicolumn{1}{c}{$xq^{(-)}$} & \multicolumn{1}{c|}{$xg$}\\
\hline
$10^{-5}$ & $1.8328\cdot10^{-2}$ & $1.6522\cdot10^{-2}$ & $1.6068\cdot10^{+2}$\\
 & $-0.0244\cdot10^{-2}$ & $-0.0682\cdot10^{-2}$ & $+0.0002\cdot10^{+2}$\\
\hline
$10^{-4}$ & $4.2274\cdot10^{-2}$ & $4.0850\cdot10^{-2}$ & $7.1187\cdot10^{+1}$\\
 & $-0.0834\cdot10^{-2}$ & $-0.1367\cdot10^{-2}$ & $+0.0007\cdot10^{+1}$\\
\hline
$10^{-3}$ & $1.1500\cdot10^{-1}$ & $1.1325\cdot10^{-1}$ & $2.6591\cdot10^{+1}$\\
 & $-0.0337\cdot10^{-1}$ & $-0.0283\cdot10^{-1}$ & $+0.0002\cdot10^{+1}$\\
\hline
$10^{-2}$ & $3.4757\cdot10^{-1}$ & $3.4617\cdot10^{-1}$ & $7.5377\cdot10^{+0}$\\
 & $-0.1013\cdot10^{-1}$ & $-0.0734\cdot10^{-1}$ & $+0.0002\cdot10^{+0}$\\
\hline
$0.1$ & $7.6366\cdot10^{-1}$ & $7.8369\cdot10^{-1}$ & $9.8748\cdot10^{-1}$\\
 & $+0.0752\cdot10^{-1}$ & $+0.0597\cdot10^{-1}$ & $-0.0006\cdot10^{-1}$\\
\hline
$0.3$ & $4.4859\cdot10^{-1}$ & $4.6170\cdot10^{-1}$ & $1.0210\cdot10^{-1}$\\
 & $+0.2826\cdot10^{-1}$ & $+0.2512\cdot10^{-1}$ & $-0.0001\cdot10^{-1}$\\
\hline
$0.5$ & $1.3366\cdot10^{-1}$ & $1.3449\cdot10^{-1}$ & $9.6596\cdot10^{-3}$\\
 & $+0.1428\cdot10^{-1}$ & $+0.1263\cdot10^{-1}$ & $-0.0013\cdot10^{-3}$\\
\hline
$0.7$ & $1.7155\cdot10^{-2}$ & $1.6402\cdot10^{-2}$ & $3.8338\cdot10^{-4}$\\
 & $+0.2657\cdot10^{-2}$ & $+0.2254\cdot10^{-2}$ & $-0.0006\cdot10^{-4}$\\
\hline
$0.9$ & $1.6862\cdot10^{-4}$ & $1.4104\cdot10^{-4}$ & $5.8941\cdot10^{-7}$\\
 & $+0.3932\cdot10^{-4}$ & $+0.2942\cdot10^{-4}$ & $-0.0011\cdot10^{-7}$\\
\hline

\end{tabular}
\end{center}
\end{footnotesize}
\caption{Comparison between the exact and the truncated method of \textsc{Candia} at $\mu_F=\mu_R=100\,\mbox{GeV}$ in the variable flavor number scheme. In each entry, the first number is the result from the exact method and the second one is the difference between the exact and the truncated method. The columns for which the exact and the truncated methods give the same results are not shown. The MRST parametrizations at $1\,\mbox{GeV}$ have been used as initial conditions.}
\label{tab:EXvsTR}
\end{table}

\clearpage

\section{Conclusions}

We have presented a new evolution program, \textsc{Candia}, which solves the DGLAP equation with high precision and 
is completely implemented in $x$-space. We have also briefly discussed the types of solutions which are implemented in the program, 
from the truncated to the exact ones. In the non-singlet sector we have shown how to construct the exact solutions analytically 
by the iteration of expressions which resum the simple logarithms of the ratio of the two couplings at the initial and final evolution 
scales. We have also addressed the issue of accuracy of the solutions and illustrated the difference between {\em brute force} 
and analytical methods, showing the connection between our approach and more traditional approaches based on the inversion of Mellin 
moments. We hope to return in the near future with additional numerical implementations, which will
provide the user with all the necessary tools so to proceed with an independent partonometric
analysis of the LHC data on the PDFs.

\section*{Acknowledgements}
\addcontentsline{toc}{section}{Acknowledgements}

M.G.~thanks the 
Theory Group at the Univ. of Liverpool or hospitality and partial 
financial support. The work of C.C.~was supported (in part)
by the European Union through the Marie Curie Research and Training Network
``Universenet'' (MRTN-CT-2006-035863) and by The Interreg II Crete-Cyprus 
Program. He thanks the Theory group at Crete for hospitality. The work of A.C.~is supported by the Transfer of Knowledge program \textsc{Algotools} (MTKD-CT-2004-014319).


\begin{thebibliography}{10}

\bibitem{Vogt:2004ns}
A. Vogt,
\newblock Comput. Phys. Commun. 170 (2005) 65, hep-ph/0408244.

\bibitem{QCDNUM}
M. Botje,
\newblock http://www.nikhef.nl/~h24/qcdnum/

\bibitem{HOPPET}
G. Salam,
\newblock http://projects.hepforge.org/hoppet/

\bibitem{Cafarella:2005zj}
A. Cafarella, C. Corian\`{o} and M. Guzzi,
\newblock Nucl. Phys. B748 (2006) 253, hep-ph/0512358.

\bibitem{Ellis:1993rb}
R.K. Ellis, Z. Kunszt and E.M. Levin,
\newblock Nucl. Phys.  B420 (1994) 517.

\bibitem{Cafarella:2007tj}
A. Cafarella, C. Corian\`{o} and M. Guzzi,
\newblock JHEP 08 (2007) 030, hep-ph/0702244.

\bibitem{Cafarella:2003jr}
A. Cafarella and C. Corian\`{o},
\newblock Comput. Phys. Commun. 160 (2004) 213, hep-ph/0311313.

\bibitem{Chetyrkin:1997sg}
K.G. Chetyrkin, B.A. Kniehl and M. Steinhauser,
\newblock Phys. Rev. Lett. 79 (1997) 2184, hep-ph/9706430.

\bibitem{Buza:1996wv}
M. Buza et~al.,
\newblock Eur. Phys. J. C1 (1998) 301, hep-ph/9612398.

\bibitem{Moch:2004pa}
S. Moch, J.A.M. Vermaseren and A. Vogt,
\newblock Nucl. Phys. B688 (2004) 101, hep-ph/0403192.

\bibitem{Vogt:2004mw}
A. Vogt, S. Moch and J.A.M. Vermaseren,
\newblock Nucl. Phys. B691 (2004) 129, hep-ph/0404111.

\bibitem{Gehrmann:2001pz}
T. Gehrmann and E. Remiddi,
\newblock Comput. Phys. Commun. 141 (2001) 296, hep-ph/0107173.

\bibitem{Remiddi:1999ew}
E. Remiddi and J.A.M. Vermaseren,
\newblock Int. J. Mod. Phys. A15 (2000) 725, hep-ph/9905237.

\bibitem{Martin:2001es}
A.D. Martin et~al.,
\newblock Eur. Phys. J. C23 (2002) 73, hep-ph/0110215.

\bibitem{Martin:2002dr}
A.D. Martin et~al.,
\newblock Phys. Lett. B531 (2002) 216, hep-ph/0201127.

\bibitem{Alekhin:2002fv}
S. Alekhin,
\newblock Phys. Rev. D68 (2003) 014002, hep-ph/0211096.

\bibitem{press}
W. Press et~al.,
\newblock Numerical Recipes in C, 2nd ed. (Cambridge University Press,
  Cambridge, UK, 1992).

\bibitem{Giele:2002hx}
W. Giele et~al.,
\newblock (2002), hep-ph/0204316.

\bibitem{Furmanski:1981ja}
W. Furmanski and R. Petronzio,
\newblock Nucl. Phys. B195 (1982) 237.

\end{thebibliography}

\end{document}